\documentclass[aps,pre,twocolumn,superscriptaddress,floatfix,10pt]{revtex4-1}   

\usepackage[version=4]{mhchem} 
\usepackage{braket}
\usepackage{siunitx}
\usepackage{amsmath}
\usepackage{floatrow}
\usepackage{graphicx}
\usepackage{hyperref}
\usepackage{gensymb}
\usepackage{datetime}
\usepackage{enumitem}
\usepackage{bm}
\usepackage{dcolumn}
\usepackage{xcolor}

\definecolor{mRed}{RGB}{0,0,0}
\newcommand{\newtext}[1]{\textcolor{mRed}{#1}}

\begin{document}

\preprint{APS/123-QED}

\title{Accurate prediction of macroscopic transport from microscopic imaging via critical fractals at the Mott transition}

\author{P.-Y. Chen}
\affiliation{Department of Physics and Astronomy, Purdue University, West Lafayette, IN 47907, USA}
\affiliation{Purdue Quantum Science and Engineering Institute, West Lafayette, IN 47907, USA}
\author{A.~R.~Rajapurohita}
\affiliation{Department of Physics and Astronomy, Purdue University, West Lafayette, IN 47907, USA}
\affiliation{Purdue Quantum Science and Engineering Institute, West Lafayette, IN 47907, USA}
\author{M.~Alzate Banguero}
\affiliation{Laboratoire de Physique et d'\'Etude des Mat\'eriaux, ESPCI Paris, PSL Université, CNRS, Sorbonne Université, 75005 Paris, France}
\author{S.~Basak}
\affiliation{Department of Physics and Astronomy, Purdue University, West Lafayette, IN 47907, USA}
\affiliation{Purdue Quantum Science and Engineering Institute, West Lafayette, IN 47907, USA}
\author{F.~Simmons}
\affiliation{Department of Physics and Astronomy, Purdue University, West Lafayette, IN 47907, USA}
\affiliation{Purdue Quantum Science and Engineering Institute, West Lafayette, IN 47907, USA}
\author{P.~Salev}
\affiliation{Department of Physics and Center for Advanced Nanoscience, University of California-San Diego, La Jolla, California 92093, USA}
\author{L.~Aigouy}
\affiliation{Laboratoire de Physique et d'\'Etude des Mat\'eriaux, ESPCI Paris, PSL Université, CNRS, Sorbonne Université, 75005 Paris, France}
\author{Ivan.~K.~Schuller}
\affiliation{Department of Physics and Center for Advanced Nanoscience, University of California-San Diego, La Jolla, California 92093, USA}
\author{A.~Zimmers}
\affiliation{Laboratoire de Physique et d'\'Etude des Mat\'eriaux, ESPCI Paris, PSL Université, CNRS, Sorbonne Université, 75005 Paris, France}
\author{E.~W.~Carlson}
\affiliation{Department of Physics and Astronomy, Purdue University, West Lafayette, IN 47907, USA}
\affiliation{Purdue Quantum Science and Engineering Institute, West Lafayette, IN 47907, USA}

\date{\today}

\begin{abstract}
Vanadium dioxide (\ce{VO2}) exhibits hysteresis in resistance while undergoing a thermally driven insulator-metal transition (IMT). Understanding the nonequilibrium effects in resistance is of great interest, as \ce{VO2} is a strong candidate for brain-inspired computing, which is more energy efficient for AI tasks compared to traditional computing. Accurate models of the connection between microscopic and macroscopic transport properties and microscopic imaging of \ce{VO2} will allow us to better utilize \ce{VO2} in future applications. However, predictions of macroscopic resistance of \ce{VO2} that quantitatively match observations using spatially resolved data have not yet been achieved.  Here, we demonstrate an accurate prediction of the macroscopic resistance of \ce{VO2} throughout the entire temperature range of interest,  by developing a multiscale resistor network model incorporating the assumption of fractal sub-pixel structure of the optical data, where the configuration of insulating and metallic domains within each pixel are drawn from the random field Ising model near criticality.  This strongly indicates that the observed fractal, power law structure of metallic and insulating domains extends down to much smaller length scales than the current record for experimental resolution of this system, and that the two-dimensional random field Ising model near criticality is a suitable model for describing the metal and insulator patches of \ce{VO2} down to scales that approach the unit cell. 
\end{abstract}
\maketitle


\section{\label{sec:introduction}Introduction}

In the search for exotic phases in quantum materials and new applications, one of the most common characterization methods after synthesizing a new sample is macroscopic transport. In many quantum materials, however, electronic phase separation is present, complicating the analysis of these measurements~\cite{Fath1999, Mathur2003, Dagotto2005, McElroy2005, Qazilbash2007, Post2018, Shi2020, Song2023}. To address this problem, various probes have reported metallic and insulator surface/volume fractions over the years. Some used bulk properties, such as magnetization~\cite{Becker2002} or NMR intensity in manganites~\cite{Dho1999}, while others used surface probes like scanning tunneling  microscopy (STM) in manganites ~\cite{Becker2002} and high-Tc superconductors~\cite{McElroy2005}, and scanning near-field optical microscopy (SNOM)~\cite{Qazilbash2011} or X-ray diffraction~\cite{Kumar2014} in \ce{VO2}. These volume fractions
have then been used as a key ingredient to calculate the macroscopic resistance using models such as effective medium approximation or resistor networks. While resistor networks have come the closest to accurately modeling these phenomena, they have struggled to quantitatively capture the temperature dependence
of the IMT in transition metal oxides such as 
\ce{La_{1-x}Sr_xMnO_3}, \ce{VO2},\ce{V2O3}, \ce{NdNiO3}, and \ce{SmNiO3},
typically {\em off by an order of magnitude} or more
in certain temperature 
ranges.\footnote{See {\em e.g.}, Fig.~4 of Ref.~\cite{Becker2002}, Fig. 2 of Ref.~\cite{Rocco2022}, Fig. 4(b) of Ref.~\cite{rozenberg-RN-nickelates}, Fig. 4 of Ref.~\cite{Wang2015}, Fig. 4 of Ref.~\cite{Guenon2013}, and Figs. 4 and 10 of Ref.~\cite{Lange2021}.
}
The issue is that transport predicted from the electronic inhomogeneity
results in overly abrupt changes in the macroscopic resistance 
during the insulator-metal phase transition compared to the experimentally measured macroscopic resistance $R(T)$. This limits any precise comparison of macroscopic system properties (for example memory in \cite{ramp-reversal}) 
to microscopic insulating fraction.
Gu\'enon {\em et al.} reached a very close match with a phenomenological model, 
but did not explicitly connect microscopic measurements to macroscopic results.\cite{Guenon2013}
Indeed, the resistance hysteresis predicted by metal-insulator domains depends not just on the insulating fraction, but also depends sensitively on the morphology of those domains, whether they
are set by, {\em e.g.}, a simple non-interaction percolation model
or whether interactions are important. Here, we fully take into account the morphology of domains measured optically into a resistor network model and demonstrate that the fractal structure of the domains
due to proximity to the critical end point of the coexistence region must be accounted for to resolve the discrepancies.

\newtext{
As \ce{VO2} goes through the metal-insulator transition, both resistance avalanches\cite{sharoni-avalanches} and surface imaging probes like SNOM\cite{shuo-prl} and optical imaging\cite{basak-deep-learning}
display evidence of this criticality.
When a system is near criticality, power law behavior appears in every measure, including in this case the metal and insulator domain morphology, which gives rise to the fractal electronic structure seen in \ce{VO2} as it transitions.\cite{shuo-prl,basak-deep-learning,Qazilbash2007}
For any system near criticality, the associated power law behavior extends from the short distance cutoff (the atomic unit cell size) all the way up to a large distance cutoff set by the smaller of the correlation length or the system size.  The correlation length itself diverges as criticality is approached. 
Because the short distance cutoff of the power laws is smaller than the resolution of the SNOM and optical imaging, when images derived from these display fractal textures, critical theory predicts that the same fractal behavior is also happening {\em inside of each pixel.}  In this paper, we use simulations near criticality to predict the fractal structure inside of each pixel, and thereby achieve a {\em quantitative} match between the optical images and macroscopic resistivity.
}

To do so, we have used a series of autofocused, auto-aligned images of the surface of \ce{VO2}, reported recently~\cite{AlzateBanguero2025}, to track the metal-insulator patches down to optical wavelength resolution throughout the transition while simultaneously gathering macroscopic transport data. These images carry not only spatial information of cluster formation but also grayscale information {\em within each pixel}. Combining this optical data with a random field Ising model near criticality and 2D resistor network gives us a unique opportunity to simulate, with minimal free parameters, the macroscopic resistance throughout the metal-insulator transition and capture fine changes even near the threshold at which domains first start to percolate. Our results enable macroscopic resistance to be predicted accurately (within the resistance’s standard error bar) from only optical measurements. This could help estimate the local resistance of nano-textured materials when electrical contacting is impractical \cite{Kepi2025} and, in some cases, reduce the need for complex, time-consuming nanoscale imaging without sacrificing the precision of macroscopic predictions.

\section{\label{sec:rationale}Rationale}

Due to the high conductivity contrast between the metal and insulating patches, several groups have used local surface images over the years to try and predict macroscopic resistance IMT in transition metal oxides.  
To do so, a series of surface images must first be recorded and then used as inputs in a resistor network or effective medium simulation. To record such images, local probe measuring campaigns are 6 to 28 scans \cite{Qazilbash2007, Qazilbash2011, Liu2013, Stinson2018, Spitzig2022}, due to time and stability (sample and tip) limitations. To address this, optical mapping of temperature and electrically triggered transitions has been reported, as it is full-field and does not require scanning. However, due to defocusing during temperature sweeps, most teams have been limited to reporting electrically triggered transitions optically (where the sample remains in focus).

In all cases, the general shape of the simulations of the IMT  transition is qualitatively similar to the measured macroscopic resistance, but a true quantitative match was not achieved throughout the temperature range of interest, with most comparisons being off by an order of magnitude or more in certain temperature ranges.  
For example, Ref.~\cite{Becker2002} used scanning tunneling spectroscopy to extract the local tunneling conductance, then used that as input for a resistor network model.  However, due to the limitations of a scanning probe, only a small part of the sample was measured via STS, and the comparison deviates at high temperature (Fig.~4 of Ref.~\cite{Becker2002}).    
Wang {\em et al.} use a non-interacting percolation model to predict 
the resistance curve of \ce{V2O3} throughout the entire hysteresis region (Fig.~4 of Ref.~\cite{Wang2015}), however, the prediction for the insulating and metallic regions changes too slowly while the transition phase changes too drastically compared to the experimental data.\cite{Wang2015} 
Filamentary effects under conditions of applied voltage have been qualitatively explained
through a model by which thermally activated two-state switchers interact
via Joule heating.\cite{Rocco2022,Stoliar2013}  However, the comparisons are qualitative rather than quantitative, and can be off by orders of  magnitude in certain ranges ({\em e.g.}, Fig.~2 of Ref.~\cite{Rocco2022}).    
One auto-focused optical study was able to closely match the electrically driven breakdown transport curves using a resistor network (Fig.~3 of Ref.~\cite{Lange2021}). For the temperature-dependent $R(T)$ simulation, only the volume fraction \( f \) was adjusted in an effective medium calculation.  However, the  drop in resistance is too abrupt as it approaches the metallic state (Fig.~10 of Ref.~\cite{Lange2021}).
A remarkably favorable comparison was achieved by Guénon {\em et al.} on \ce{V2O3} using a model of independent hysteretic switchers ({\em i.e.}, the Preisach model from magnetism)\cite{Preisach1935},  that interact only through local Joule heating.\cite{Guenon2013}  However, their phenomenological model  did not explicitly connect microscopic imaging data to the macroscopic result.

Indeed, achieving a true quantitative comparison by which theory can
successfully connect the microscopic imaging with  macroscopic transport 
throughout the hysteretic temperature range has 
thus far remained elusive.  
In what follows below, we first show that optical measurements can produce sufficiently detailed spatially resolved images of the IMT at small temperature increments in VO\(_2\). We then use this series of images to test a standard resistor network,
which is also off by an order of magnitude in certain temperature regions.  
Finally, in Secs.~\ref{sec:need-subpixel}-\ref{sec:full-model}  we show 
that a sub-pixel fractal resistor network can {\em quantitatively} 
model experimental data on macroscopic transport using microscopic imaging as inputs. 
\newtext{This is consistent with the expectation from critical theory that the fractal nature of the domains observed in SNOM and optical data extends inside of each pixel as well. }


\section{Experimental Methods}
\label{sec:experiment}
\begin{figure}[]
    \centering
    \includegraphics[width=0.9\textwidth]{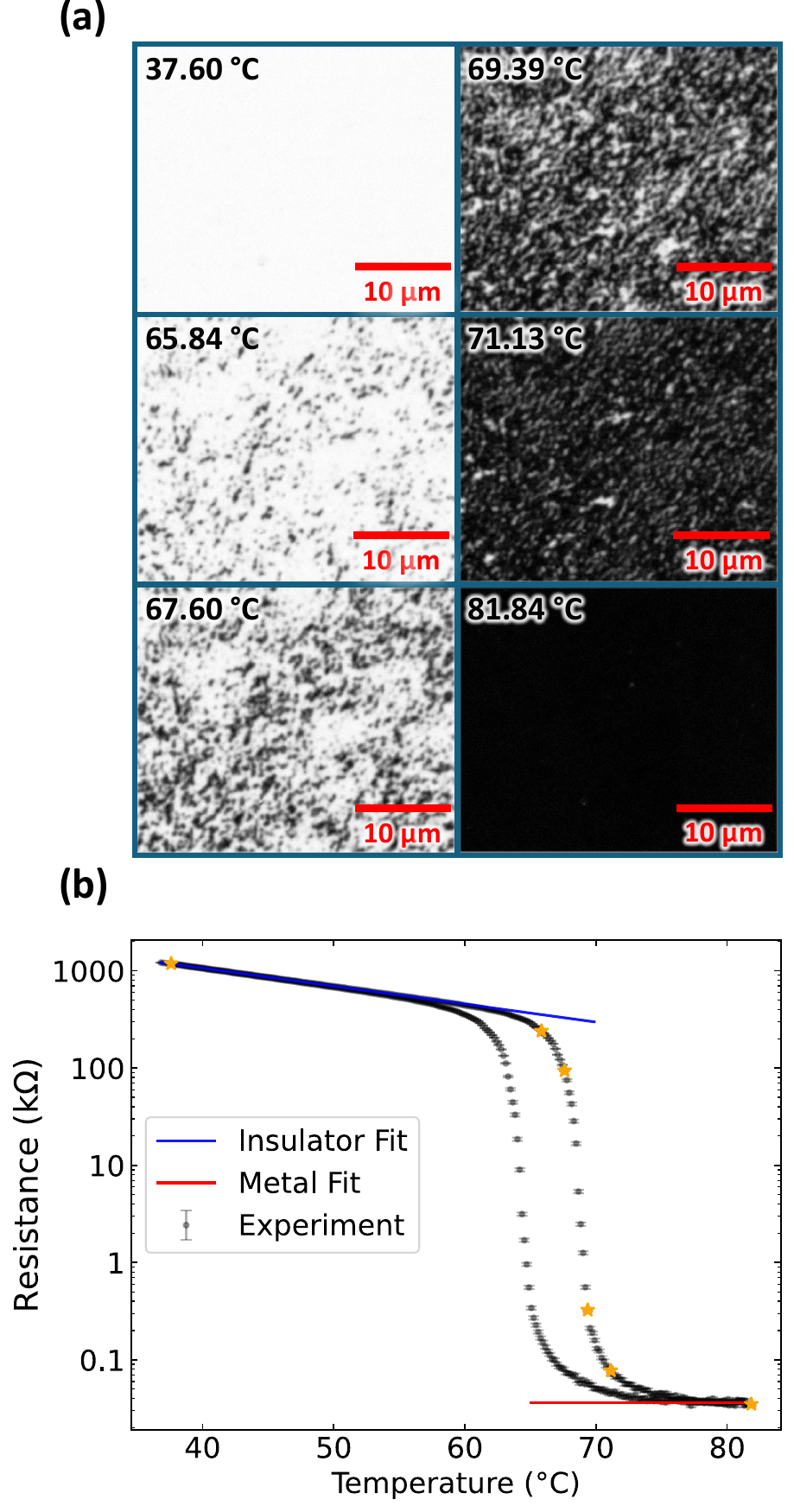}
    \caption{Experimental data consisting of: (a) Grayscale optical microscopy images of \ce{VO2} at various temperatures. The intensity of each pixel has been normalized for consistency. (b) Experimental data of macroscopic resistance versus temperature. The blue and red lines are linear fits on the log-linear scale where the blue line is the insulating fit and the red line is the metallic fit. The insulating region ($T\leq 40$\textdegree C) is fit to an exponential activation $R_{\text{insulating}}(T) = R^0_{\text{insulating}}e^{-AT}$. The metallic region ($T\geq 80$\textdegree C) is fit to a constant $R_{\text{metallic}}(T) = R^{0}_{\text{metallic}}$. The orange stars represent the corresponding position on the $R(T)$ curve for the images in panel (a).
    }
    \label{fig:grayscale_transition_and_fit}
\end{figure}

A \ce{VO2} thin film, 100 nm thick, was grown on an r-cut sapphire substrate using RF magnetron sputtering~\cite{PhysRevB.79.235110}. The optical experimental setup involved positioning the thin film on a Linkam Thms350V temperature controller, placed within a Nikon optical microscope in an epi configuration (where both illumination and light reflection pass through the same objective). Visible light illumination was provided by a halogen lamp without filters. Images of the sample’s surface were captured near the focal point, approximately 1 mm, using a $\times$150 magnification dry Olympus objective lens with an optical aperture of $NA = 0.9$. The field of view of the images was 33 microns $\times$ 28 microns, which is 910 $\times$ 760 pixels. The theoretical lateral resolution, calculated using the Rayleigh criterion, is $\delta r = 1.22\lambda/(2NA)$ = 370 nm in the visible range. Image processing involved correcting for illumination inconsistencies, temperature drift, and alignment, as well as single pixel scaling to obtain the grayscale images used as input for the model~\cite{AlzateBanguero2025}. The sample exhibited a 27\% relative optical change in the visible range during the IMT transition. Resistance was recorded for each frame during the imaging process, see Fig.~\ref{fig:grayscale_transition_and_fit}(b). 

Using a fully aligned series of images, each pixel is normalized from fully white to fully black (see Fig.~\ref{fig:grayscale_transition_and_fit}(a)), during the insulating-to-metallic transition. This grayscale optimization is similar to standard image enhancement techniques used in daily life, except that here the normalization is performed for each pixel independently. A full description of this pixel-by-pixel normalization method is provided in~\cite{AlzateBanguero2025}. In Fig.~\ref{fig:grayscale_transition_and_fit}(a), the full white/black images correspond to an optical image taken during the insulating/metallic transition. Upon warming from a fully insulating state, small metallic domains nucleate (black spots) within the full white (insulating state) image, but as $T$ keeps increasing more metallic domains nucleate, while the ones that exist continue to grow and even combine with other metallic patches eventually leading to the image being fully black (fully metallic).  
The pattern of metal and insulator regions observed in the optical data is reminiscent of the fractal textures that arise near criticality of the random field Ising model\cite{basak-deep-learning,shuo-prl}, and also reminiscent of the textures seen in the more microscopic treatment of Ref.~\cite{vlad-RF-RN}.

As the transition progresses, the changes in the optical image also correspond to resistance jumps in the resistance curve $R(T)$ shown in Fig.~\ref{fig:grayscale_transition_and_fit}(b).
The resistance jumps are sensitive not just to the number of pixels that switch,
but also to the shape of the avalanche and how it connects to other regions, necessitating the detailed simulations we provide in this paper.  
Our goal is to establish a theoretical model to connect the microscopic spatial information from the optical microscope to the macroscopic resistance measurement of the \ce{VO2} film.

\section{Theoretical Model and Results}
\label{sec:methods}

\subsection{Basic Resistor Network}
\label{sec:resistor_network}

Our goal is to develop a method by which macroscopic transport in the mixed phase of 
the Mott transition can be 
{\em accurately predicted from surface imaging experiments}.
We begin by describing the most basic resistor network model one can draw from 
optical data. As we will see, this model is inadequate, and
will be superseded by the model we develop in 
Secs.~\ref{sec:need-subpixel}-\ref{sec:full-model} 
below.  There, we develop a multiscale resistor network method,
which assumes that the power law, fractal structure seen in
images such as Fig.~\ref{fig:grayscale_transition_and_fit}(a) also extends into the sub-pixel regime. 

\begin{figure}[]
    \includegraphics[width=\textwidth]{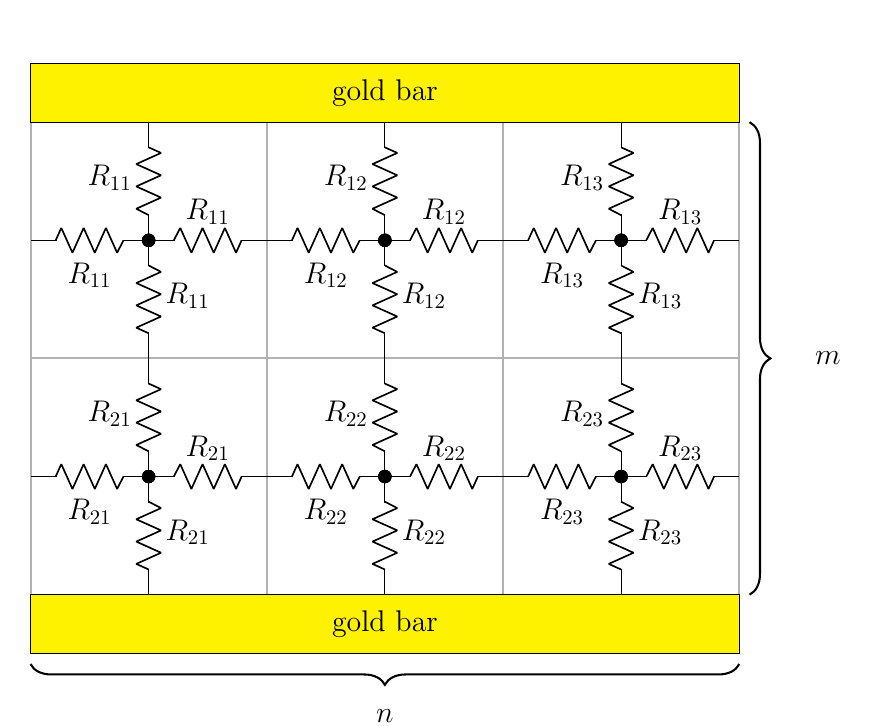}
    \caption{Each plaquette in the figure represents a single pixel 
    from an optical microscopy image. 
    Each pixel is modeled as four resistors coming out of a node (solid black circle),
    with the value of all four resistors set by the grayscale value of the pixel.  In this way, 
    each image is mapped to a unique resistor network.  
    We use $n$ to denote the number of plaquettes parallel to the gold bus bars,
    and $m$ to denote the number of plaquettes perpendicular to the gold pads.
    A representative (small) size $n\times m = 3 \times 2$ is shown. }
    \label{fig:resistor_network}
\end{figure}

Using optical images of the \ce{VO2} film, we model each pixel as four resistors coming out of a node, as shown in Fig.~\ref{fig:resistor_network}. In doing so we convert each image into a resistor network, which allows us to find the resistance of the \ce{VO2} film at that point in time. To find the macroscopic resistance of the entire network, we use the bond-propagation algorithm, an exact algorithm whose computational time scales as ${\cal{O}} (N^{3/2})$ where $N$ is the number of plaquettes~\cite{bp-algo}. 
By repeatedly applying the standard Y-$\Delta$ transformation 
to reduce resistor networks, 
the bond-propagation algorithm successively eliminates resistors from the network, leaving behind a single effective resistor. 

\subsection{\label{sec:temperature}Limiting cases at high and low temperatures}

To use the resistor network model to predict macroscopic resistance, we must first assign values of the resistors based on the optical data.
To do this, we first map the limiting cases, {\em i.e.}, the fully insulating state at low temperature and the fully metallic state at high temperature.
In these cases, the microscopic resistance $(R_{\text{micro}})$ for each of these resistors is related to the geometry of the film and the macroscopic resistance $(R_{\text{macro}})$ as follows:
\begin{align}\label{eqn:geometry}
    R_{\text{macro}} &= \frac{2mR_{\text{micro}}}{n}
\end{align}
where $n,m$ are defined in Fig.~\ref{fig:resistor_network}. To find $R(T)$ in these two regions, we model the insulating resistance using an exponential activation (blue line) in temperature while the metallic resistance is modeled as a constant (red line), as shown in Fig.~\ref{fig:grayscale_transition_and_fit}(b). 
In the insulating region, we let $R_{\text{macro}} = R_{\text{insulating}}(T)$ to find the microscopic resistance $R_{\text{micro}}^{\rm insulating}(T)$ in the insulating region.
In the metallic region, we let $R_{\text{macro}} = R_{\text{metallic}}^0$ to find the microscopic resistance $R_{\text{micro}}^{\rm metallic}$ in the metallic region. The exponential activation for the insulating resistance is $R_{\text{insulating}}(T) = R^0_{\text{insulating}}e^{-AT}$ where the parameters have values $R^{0}_{\text{insulating}} = 5.83\cdot 10^6$ $\Omega$ and $A = -0.043(\text{\textdegree C})^{-1}$. The value of the fit for the constant resistance in the metallic state is $R^{0}_{\text{metallic}} = 36.38$ $\Omega$.

\subsection{Resistor Network: The Need for Fractal Sub-Pixel Structure}
\label{sec:need-subpixel}

With the limiting cases in hand ({\em i.e.}, for the fully insulating and fully metallic resistances) we can now address the transition process. The most direct approach is to use black-and-white images obtained with a threshold (as in Ref.~\cite{AlzateBanguero2025}) from Fig.~\ref{fig:grayscale_transition_and_fit}(a) as input to the resistor network. The resulting $R(T)$ simulated curve, shown in Fig. \ref{fig:combined_results}(a), exhibits severe shortcomings, specifically a transition that is too steep, as reported in previous studies \cite{Wang2015, Guenon2013, Lange2021}. Thus, a simple binary resistor network is not sufficient to capture the resistance properties during the IMT.  
To overcome this, 
we next develop a multiscale method to 
take advantage of the information in the grayscale of each pixel to better estimate the resistance of each node in the resistor network.

As mentioned in Sec.~\ref{sec:rationale}, we previously showed that the two-dimensional random field Ising model (2D RFIM) best describes the critical behavior of \ce{VO2}. Critical cluster techniques developed in our group demonstrated that the 2D RFIM best accounts for the power-law structure of metallic and insulating domains during the Mott transition in \ce{VO2} thin films~\cite{shuo-prl}. Furthermore, the power-law behavior of avalanches and clusters extends over multiple decades of scaling within the resolution of both optical and SNOM methods and is not cut off, suggesting that this behavior persists beyond the resolution of the optical probe. This provides insight into the range of domain sizes that should be tested in our model.
More recently, we trained a deep learning convolutional neural network to recognize the minimal model driving electronic pattern formation in quantum materials, and applied it to our optical microscopy data on a thin film of \ce{VO2}. Our deep learning model also identified the 2D RFIM as the primary driver of the spatial correlations of the metal and insulator domains as \ce{VO2} films transition~\cite{basak-deep-learning}.

We can incorporate this knowledge by pairing it with the grayscale optical images of \ce{VO2} (Fig.~\ref{fig:grayscale_transition_and_fit}(a)). Since grayscale pixels have values between 0 (fully metallic domains) and 1 (fully insulating domains), we assume that they represent a combination of insulating and metallic domains. Although the sub-pixel structure cannot be resolved using optical data, we propose that it follows the 2D RFIM, consistent with the scaling behavior observed in both optical and SNOM data~\cite{shuo-prl, basak-deep-learning}.

\subsection{Two-Dimensional Random Field Ising Model of Sub-Pixel Fractal Structure}
 
The 2D RFIM is described as follows: 
\begin{align}
    \mathcal{H} = -J\sum_{<ij>} \sigma_i \sigma_j + \sum_i (H+h_i) \sigma_i
\end{align}
where the pseudospins $\sigma_i = \pm 1$ denote the state of each site, with $\sigma_i = -1$ representing a metallic site and $\sigma_i = 1$ representing an insulating site. The model includes nearest-neighbor interactions of strength $J$, a local random field $h_i$ at each site, and an external uniform sweeping field $H$. By varying the uniform field $H$ the system can be driven from a metallic to an insulating state, resulting in resistance avalanches~\cite{sharoni-avalanches} and the formation of fractal metal-insulator domain structures~\cite{shuo-prl} during the transition.  
\newtext{When an experimental system is changed from insulator to metal (or vice versa) via a change in temperature, the experimental temperature $T$ maps to  uniform field $H$ in our model.
}


The local random field $h_i$ is sampled from a Gaussian distribution centered at zero, with the standard deviation $\Delta$ representing the strength of the disorder in the material. 
\begin{align}
    P(h_i) = \frac{1}{\sqrt{2\pi}\Delta} e^{-h_i^2/2\Delta^2}
\end{align}
where the parameter $\Delta$ determines the width of the Gaussian distribution and hence the degree of disorder present in the material.

\begin{figure}[h]
    \centering
    \includegraphics[trim={1 0 1.5 0},clip,width=\textwidth]{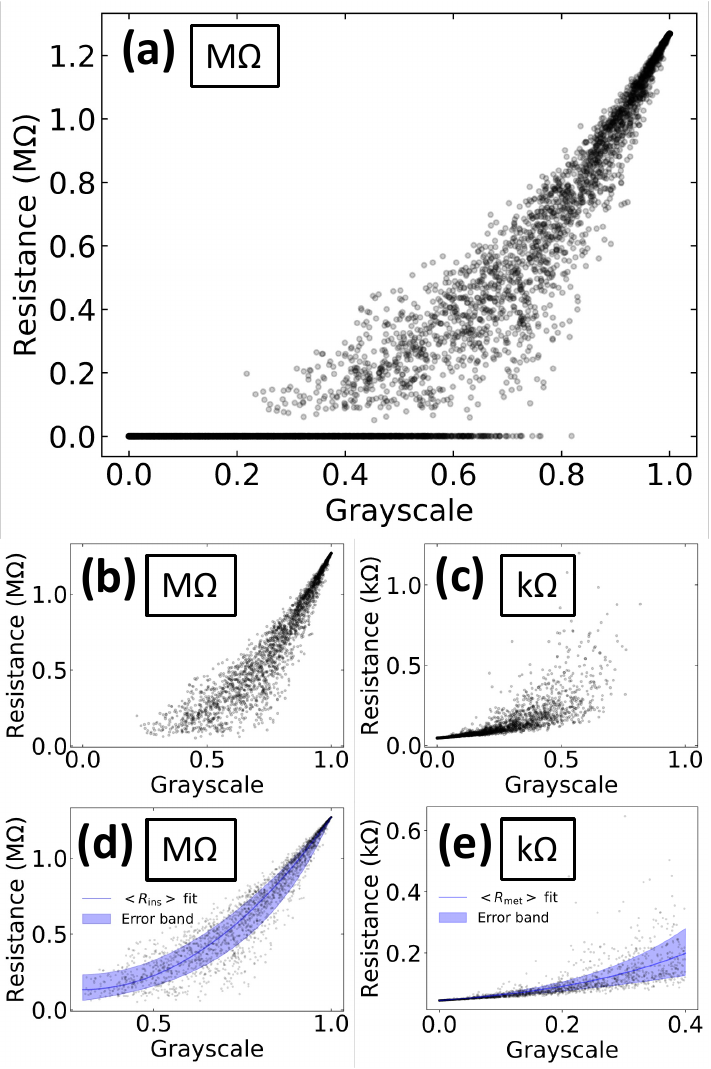}
    \caption{Mapping random field Ising configurations to macroscopic resistance.  (a) Macroscopic resistance vs. average intensity (reported as grayscale) calculated from 4608 $100\times 100$ windows of critical random field Ising configurations. Note the insulating and metallic branches. Resistance is considered insulating if over 10 k$\Omega$ (b) and metallic (c) if less than 10 k$\Omega$.
    Phenomenological fit to fractal sub-pixel structure predicted by the 2D RFIM, on the insulating branch (d) and metallic branch (e). Data points are put into bins of width 0.01 on the grayscale axis. We obtain the $\braket{R}$ fit curves by fitting to the points computed from the mean value in each bin. The blue region shows one standard deviation within each grayscale bin. Points on the insulating branch are fit down to a grayscale value of 0.3, while points on the metallic branch are fit up to a grayscale value of 0.4. The fitting functions used can be found in Table.~\ref{table:fitting-functions}, and the parameters are in Table~\ref{table:fitting-parameters-s100}.
    }
    \label{fig:sub-pixel_s100_combined}
\end{figure}

\begin{figure*}[ht]
    \centering
    \includegraphics[width=\textwidth]{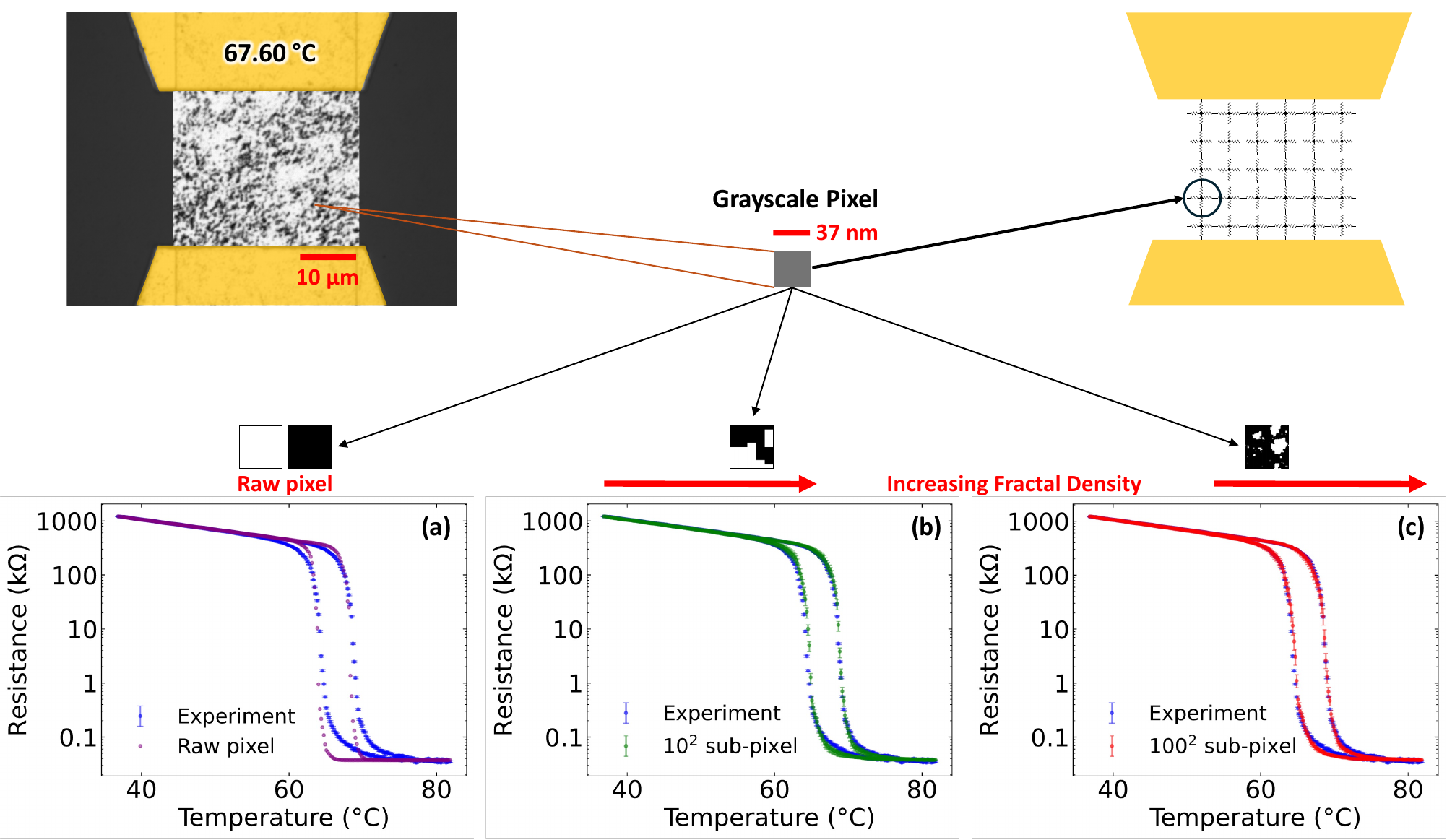}
    \caption{Schematic of the combined experiment-theory fractal resistor network procedure. Starting from the grayscale experimental surface image (top left), a resistor network was mapped out (top right). Each node (solid black circle) in the resistor network represents one pixel in the image. We compare three different ways to set the value of each resistor: (a) Starting from the experimental image, we use the DOMain INtensity Overturn (DOMINO) method we introduced in~\textcite{Basak2023} to map each pixel to either fully insulating (white pixel) or fully metallic (black pixel). (b) Resistance prediction using the $10\times 10$ fractal sub-pixel fit curve and error band in Table.~\ref{table:fitting-parameters-s10}. (c) Resistance prediction using the $100\times 100$ fractal sub-pixel fit curve and error band in Fig.~\ref{fig:sub-pixel_s100_combined}(d) and Fig.~\ref{fig:sub-pixel_s100_combined}(e). In panels (b) and (c) the grayscale threshold to switch between insulating and metallic branches was set to 0.32.}
    \label{fig:combined_results}
\end{figure*}

\subsection{Relating pixel grayscale to pixel resistance through a sub-pixel resistor network}
\label{sec:sub-pixel}

We first use a sub-pixel resistor network to predict the local
pixel resistance from the locally measured pixel grayscale of the optical measurements.
To incorporate the RFIM simulations described above into a sub-pixel resistor network, we first convert each Ising pseudospin configuration into a resistor network and use the bond-propagation algorithm to calculate the macroscopic resistance.
To do this, we map the pseudospin -1 to a metallic domain and 1 to a insulating domain. 
These pseudospins are then substituted with the microscopic resistance values, $R_{\rm micro}^{\rm insulating}(T = 40\degree\rm C) = 633423.46\,\Omega$ and $R_{\rm micro}^{\rm metallic} = 23.57\,\Omega$, as found in Sec.~\ref{sec:temperature}.
The grayscale of the grid is computed as $g = (m+1)/2$ where $m=\sum_i \sigma_i/N$ is the sum of pseudospins divided by the total number of sites $N$. The resistance is then plotted against the grayscale of each grid in Fig.~\ref{fig:sub-pixel_s100_combined}(a).
The RFIM configurations used in the sub-pixel resistor network are $100\times 100$ windows cropped from critical 2D RFIM configurations of size $10080 \times 10080$, allowing us to simulate the local behavior of a single pixel in the optical image.

For each grayscale value ($x$-axis), multiple sub-pixel RFIM black-and-white configurations are possible, each producing a different pixel resistance value ($y$-axis). 
This arises from the fact that, for example, in a 50/50 black/white pixel configuration, percolation may or may not occur depending on the arrangement of black and white sub-pixels.
This results in a cloud of possible resistance values for each grayscale level in Fig.~\ref{fig:sub-pixel_s100_combined}(a).
Practically, one must convert the cloud of points into a single curve where each grayscale corresponds a unique resistance $R$, which will later be used to convert our grayscale images into resistor networks. The data is split into an insulating branch (Fig.~\ref{fig:sub-pixel_s100_combined}(b)) for resistances greater than 10 k$\Omega$ and a metallic branch (Fig.~\ref{fig:sub-pixel_s100_combined}(c)) for resistances less than 10 k$\Omega$. 

To obtain a unique $R$ for each grayscale, a threshold was set to determine when our pixels should be considered insulating or metallic. We use a threshold value 0.32 (see SI Sec.~\ref{sec:SI_different_thresholds}) for the grayscale, optimized to best match the experimental results on macroscopic transport. 

We perform fits on the insulating and metallic branch by first putting data points in bin widths of 0.01 on the grayscale axis. For each bin, we find the mean resistance value for that grayscale bin, and the values at $\pm 1 \sigma$ of the mean.
The dark blue lines in Fig.~\ref{fig:sub-pixel_s100_combined}(d) and (e) are fits to this mean.  The light blue bands in those panels are bounded by fits to the $\pm 1 \sigma$ in each bin.
This fitting not only gives us a unique value of the resistance for each grayscale value, but also allows us to generate a standard deviation $\sigma$, necessary to compare the simulation and experimental $R(T)$ curves as shown in Fig.~\ref{fig:combined_results}(b,c).

\subsection{From microscopic image data to macroscopic transport through a large scale resistor network}
\label{sec:full-model}

Now that we have created a mapping between pixel grayscale and pixel resistance under the assumption of sub-pixel fractal electronic inhomogeneity, 
we can convert experimental grayscale optical data into a large scale resistor network to predict the macroscopic resistance for the given image. The first step is determining whether or not the pixel is insulating or metallic based on the threshold we set earlier ($g=0.32$). Next, according to Fig.~\ref{fig:sub-pixel_s100_combined}(d) and Fig.~\ref{fig:sub-pixel_s100_combined}(e) we convert the grayscale into four resistors coming out of a node with the corresponding resistance value. However, we still need to take into account the temperature dependence of the insulating branch. Thus, based on the temperature at which the image was captured, we apply the exponential temperature scaling as mentioned in Fig.~\ref{fig:grayscale_transition_and_fit}(b) for each of the microscopic resistors. Finally, we compute the resistance of the large scale resistor network with the bond-propagation algorithm.

By applying this procedure to all frames from the experimental data, we can obtain the complete $R(T)$ curves shown in Fig.~\ref{fig:combined_results}. The error bars in Fig.~\ref{fig:combined_results} were determined by computing $R(T)$ using the $\pm\sigma$ curves in Fig.~\ref{fig:sub-pixel_s100_combined}(d,e), and comparing to $R(T)$ computed with the mean resistance curve.
We modeled each pixel either as a \(10 \times 10\) grid in Fig.~\ref{fig:combined_results}(b) or a \(100 \times 100\) grid in Fig.~\ref{fig:combined_results}(c). 
These grid sizes are on the order of a few nanometers and a few angstroms, respectively. 
Each site on these finer grids is either completely insulating or completely metallic, corresponding to values of 1 and 0, respectively. 

\begin{figure}[]
    \centering
    \includegraphics[width=0.75\textwidth]{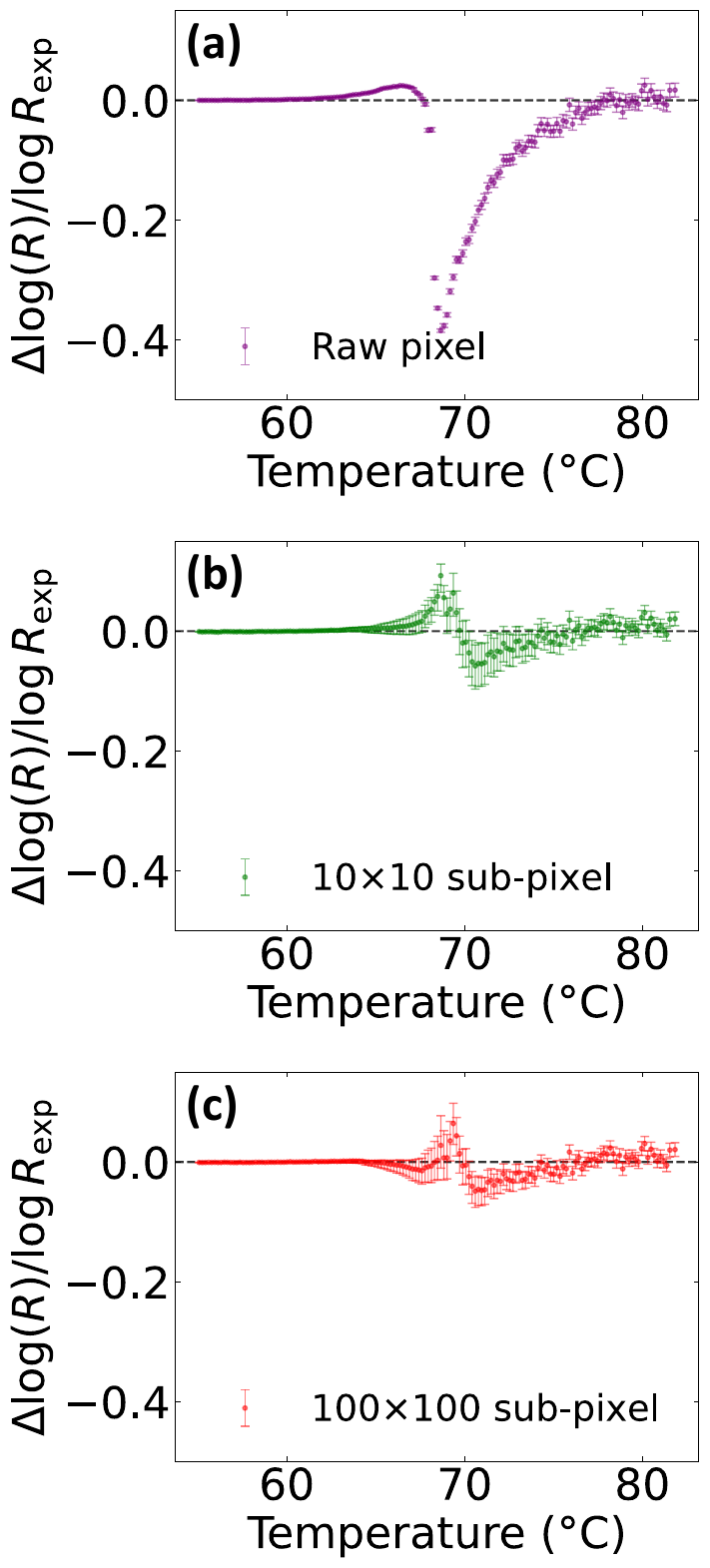}
    \caption{
    Comparison of threshold pixel and fractal RFIM sub-pixel method for macroscopic resistance hysteresis prediction from spatially resolved optical data for the warming branch. $\Delta \log R/\log R_{\rm exp}$ is plotted where $\Delta \log R = \log R_{\rm theory} - \log R_{\rm exp}$. If the error bars cross 0 (denoted by the horizontal dotted line), then the prediction matches within error bars.  
    However, large deviations between the calculated value and the horizontal dotted line indicate lack of agreement between the experimentally derived macroscopic resistance and that predicted theoretically.
    We can see that the prediction assuming sub-pixel fractal structure gives a much closer match to the data than the raw pixel prediction.  
    }
    \label{fig:delta_R_ratio}
\end{figure}

\begin{figure}[]
    \centering
    \includegraphics[width=\textwidth]{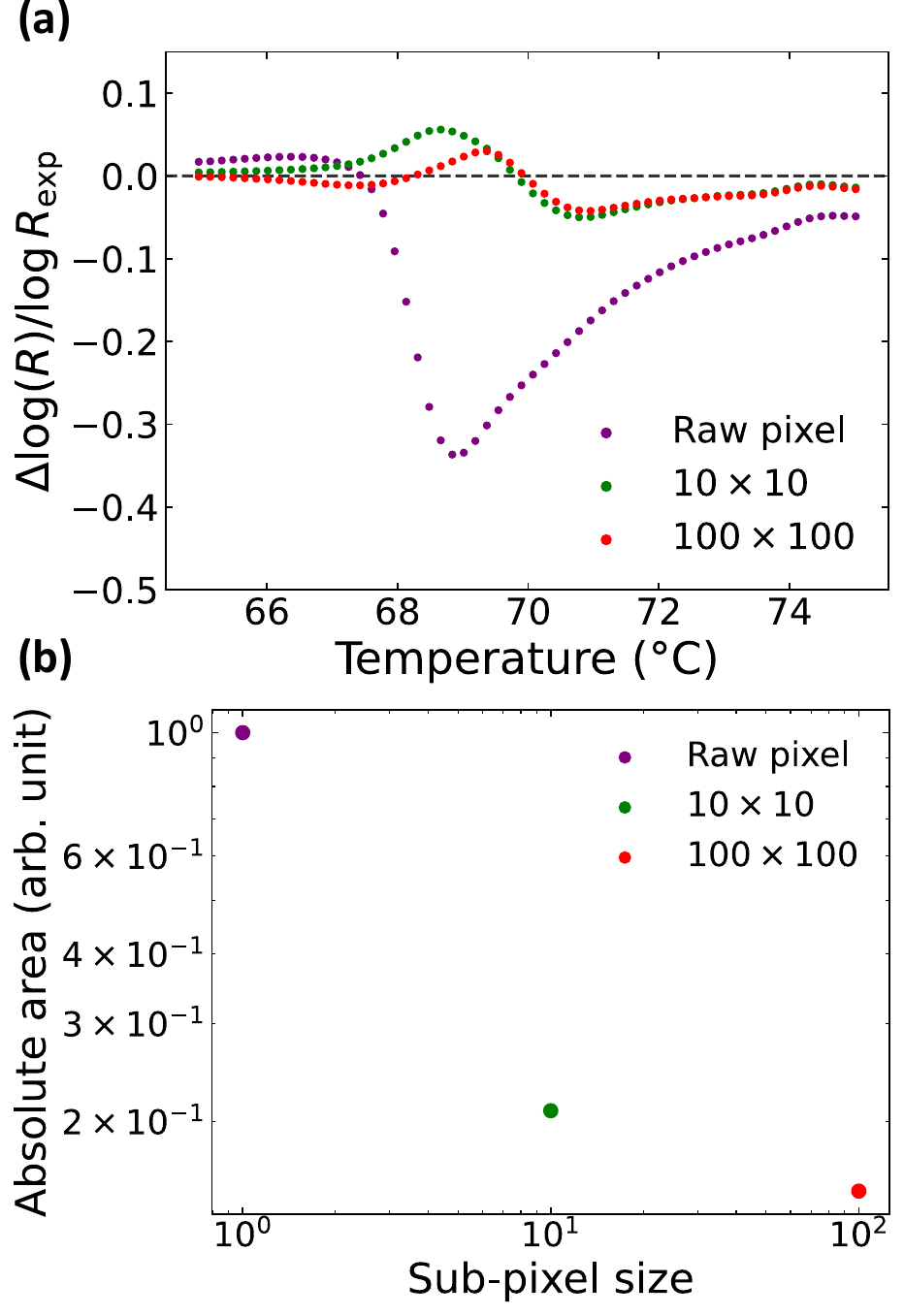}
    \caption{(a) Here we overlay the results reported in Fig.~\ref{fig:delta_R_ratio}(a,b,c), smoothed with a Gaussian of width of 0.35\textdegree C. 
    (b) By computing the absolute area under the curves in (a) we can see that the prediction becomes better as the sub-pixel grid becomes finer.
    }
    \label{fig:gaussian_combined}
\end{figure}

\section{Discussion}

Previous resistor network models used domain sizes between \(10 \times 10\) nm$^2$ and \(100 \times 100\) nm$^2$ and non-interacting models. 
For example, Ref.~\cite{Wang2015} suggested that smaller domain sizes in the resistor network would have improved their comparison to experiment. 
Here, we employ a multiscale approach.  
We simulate a large number of nodes ($910\times 760$, which is
set by the size of the optical experimental dataset), 
where each pixel is itself modeled by a smaller scale sub-pixel network of varying size %
(raw pixel, $10\times 10$, or $100\times 100$ nodes 
in the sub-pixel resistor network).\footnote{For comparison,
Guénon {\em et al.} use a resistor network of size
$20\times400$ nodes, each with 4 resistors.\cite{Guenon2013})}
\newtext{Our denser resistor networks are possible because
of our multiscale modeling approach which is warranted when
the underlying structure is fractal and the fractal structure extends down to length scales smaller than the pixel size.}

From Fig.~\ref{fig:combined_results} panels (a), (b), and (c), it is evident that our assumption of sub-pixel, fractal metal-insulator domain structure
(panels (b) and (c))
greatly enhances the comparison between theory and experiment
as compared to the raw pixel case (panel (a)).  
To quantify this comparison, in Fig.~\ref{fig:delta_R_ratio}, we plot the difference between simulated and experimental data as $\Delta \log R/\log R_{\rm exp}$ where $\Delta \log R = \log R_{\rm theory} - \log R_{\rm exp}$. 
The horizontal dashed line indicates when simulated and experimental data match exactly. 
From this comparison, it is clear that using sub-pixel of size $10\times 10$ and $100\times 100$ is significantly better than using the raw pixel prediction. 
To compare $10\times 10$ and $100\times 100$, we reduce the noise in these plots, in Fig.~\ref{fig:gaussian_combined}(a), by a gaussian smoothing. 
Then, by computing the absolute area under each curve, we obtain Fig.~\ref{fig:gaussian_combined}(b), and it is evident that using a sub-pixel of size $100\times 100$ yields better results than using a sub-pixel of size $10\times 10$, as the integrated difference with respect to zero becomes smaller for $100\times 100$.

Since our predictions for the resistance curve using a $100\times 100$ sub-pixel structure almost match all points within error bars, this gives theoretical evidence supporting that fractal structure of metallic and insulating domains appears even down to the scale of unit cell sizes.
This means that, theoretically, power-law behavior of avalanches would extend beyond the resolution of the optical microscope and down to nm (as shown by SNOM\cite{Qazilbash2011}) or even down to the scale of the unit cell size. This is hinted at in structural mapping performed by TEM \cite{Ahmadi2024, Jian2015}. If similar electronic probe resolution were achieved, future experimental work could test this prediction.

In this study, not only do we solve the problem of accurately performing resistor network simulations from microscopic measurements, but we also show that our previous understanding of the system—particularly its fractal nature—needs to be directly incorporated into the simulation. This is achieved by adding a 2D random-field Ising model at the sub-pixel level.

On a practical note, using only the grayscale information of sample surfaces together with
our sub-pixel simulations of RFIM-generated 
fractal resistor networks, 
our method enables researchers investigating the metal–insulator transition to accurately reproduce transport data when available, or to virtually determine sample resistance without the need to make contacts at the micron or nanoscale \cite{Kepi2025}.

\section{\label{sec:conclusion}Conclusion}

In summary, we have shown that the assumption of fractal sub-pixel structure 
\newtext{in the optical measurements} 
is necessary to capture the IMT of \ce{VO2}, enabling an accurate prediction of macroscopic resistance, which had previously been difficult to predict just from spatially resolved surface probes like optical data or STM. Furthermore, our work shows that for a given grayscale value, a sub-pixel assumption of $100\times 100$ is better than $10\times 10$, which is indirect evidence that the IMT is highly textured, down to length scales that approach the unit cell size of \ce{VO2},
\newtext{as expected for a system whose domain coexistence morphology is near a critical endpoint.}
Our method of combining fractal sub-pixel structure with a resistor network provides the means to accurately predict the macroscopic resistance of \ce{VO2} throughout the IMT. The methods in our study can be expanded to other materials that have fractal textures, such as NdNiO$_3$ which also has a IMT~\cite{Post2018, comin-ndnio3}, fluid flow in porous media by mapping viscosity to resistance~\cite{porous}, and also cuprate high temperature superconductors, which also host electronic fractals~\cite{Song2023}, among others. In experimental settings, our work will be useful in cases where connecting leads to internal parts to measure resistance is not feasible but optical measurements are possible. The methods described here could potentially be extended to other linear response functions like capacitance and inductance.

\section{\label{sec:acknowledgements}Acknowledgements}
We acknowledge helpful conversations with K.~A.~Dahmen and M.~D.~Golden. P.-Y.C. acknowledges support from an Open Quantum Initiative Undergraduate Fellowship, a Purdue Summer Undergraduate Research Fellowship, and a Purdue Office of Undergraduate Research Scholarship.
F.S., S.B. and E.W.C. acknowledge support from National Science Foundation Grant No.\ DMR-2006192 and the Research Corporation for Science Advancement Cottrell SEED Award. 
S.B. acknowledges support from a Bilsland Dissertation Fellowship. E.W.C. acknowledges support from a Fulbright Fellowship, and thanks the Laboratoire de Physique et d’Étude des Matériaux (LPEM) at École Supérieure de Physique et de Chimie Industrielles de la Ville de Paris (ESPCI) for hospitality. 
This research was supported in part through computational resources provided by Research Computing at Purdue, West Lafayette, Indiana.\cite{rcac-purdue}
The work at UCSD (PS, IKS) was supported by the ``Quantum Materials for Energy Efficient Neuromorphic Computing'' (Q-MEEN-C), an Energy Frontier Research Center funded by the U.S. Department of Energy, Office of Science, Basic Energy Sciences under the Award No. DE-SC0019273.
The work at ESPCI (M.A.B., L.A. and A.Z.) was supported by Cofund AI4theSciences hosted by PSL University, through the European Union’s Horizon 2020 Research and Innovation Programme under the Marie Skłodowska-Curie Grant No. 945304.


\begin{thebibliography}{39}%
\makeatletter
\providecommand \@ifxundefined [1]{%
 \@ifx{#1\undefined}
}%
\providecommand \@ifnum [1]{%
 \ifnum #1\expandafter \@firstoftwo
 \else \expandafter \@secondoftwo
 \fi
}%
\providecommand \@ifx [1]{%
 \ifx #1\expandafter \@firstoftwo
 \else \expandafter \@secondoftwo
 \fi
}%
\providecommand \natexlab [1]{#1}%
\providecommand \enquote  [1]{``#1''}%
\providecommand \bibnamefont  [1]{#1}%
\providecommand \bibfnamefont [1]{#1}%
\providecommand \citenamefont [1]{#1}%
\providecommand \href@noop [0]{\@secondoftwo}%
\providecommand \href [0]{\begingroup \@sanitize@url \@href}%
\providecommand \@href[1]{\@@startlink{#1}\@@href}%
\providecommand \@@href[1]{\endgroup#1\@@endlink}%
\providecommand \@sanitize@url [0]{\catcode `\\12\catcode `\$12\catcode `\&12\catcode `\#12\catcode `\^12\catcode `\_12\catcode `\%12\relax}%
\providecommand \@@startlink[1]{}%
\providecommand \@@endlink[0]{}%
\providecommand \url  [0]{\begingroup\@sanitize@url \@url }%
\providecommand \@url [1]{\endgroup\@href {#1}{\urlprefix }}%
\providecommand \urlprefix  [0]{URL }%
\providecommand \Eprint [0]{\href }%
\providecommand \doibase [0]{https://doi.org/}%
\providecommand \selectlanguage [0]{\@gobble}%
\providecommand \bibinfo  [0]{\@secondoftwo}%
\providecommand \bibfield  [0]{\@secondoftwo}%
\providecommand \translation [1]{[#1]}%
\providecommand \BibitemOpen [0]{}%
\providecommand \bibitemStop [0]{}%
\providecommand \bibitemNoStop [0]{.\EOS\space}%
\providecommand \EOS [0]{\spacefactor3000\relax}%
\providecommand \BibitemShut  [1]{\csname bibitem#1\endcsname}%
\let\auto@bib@innerbib\@empty
\bibitem [{\citenamefont {F{\"a}th}\ \emph {et~al.}(1999)\citenamefont {F{\"a}th}, \citenamefont {Freisem}, \citenamefont {Menovsky}, \citenamefont {Tomioka}, \citenamefont {Aarts},\ and\ \citenamefont {Mydosh}}]{Fath1999}%
  \BibitemOpen
  \bibfield  {author} {\bibinfo {author} {\bibfnamefont {M.}~\bibnamefont {F{\"a}th}}, \bibinfo {author} {\bibfnamefont {S.}~\bibnamefont {Freisem}}, \bibinfo {author} {\bibfnamefont {A.~A.}\ \bibnamefont {Menovsky}}, \bibinfo {author} {\bibfnamefont {Y.}~\bibnamefont {Tomioka}}, \bibinfo {author} {\bibfnamefont {J.}~\bibnamefont {Aarts}},\ and\ \bibinfo {author} {\bibfnamefont {J.~A.}\ \bibnamefont {Mydosh}},\ }\href {https://www.science.org/doi/abs/10.1126/science.285.5433.1540} {\bibfield  {journal} {\bibinfo  {journal} {Science}\ }\textbf {\bibinfo {volume} {285}},\ \bibinfo {pages} {1540} (\bibinfo {year} {1999})}\BibitemShut {NoStop}%
\bibitem [{\citenamefont {Mathur}\ and\ \citenamefont {Littlewood}(2003)}]{Mathur2003}%
  \BibitemOpen
  \bibfield  {author} {\bibinfo {author} {\bibfnamefont {N.}~\bibnamefont {Mathur}}\ and\ \bibinfo {author} {\bibfnamefont {P.}~\bibnamefont {Littlewood}},\ }\href {https://doi.org/10.1063/1.1554133} {\bibfield  {journal} {\bibinfo  {journal} {Physics Today}\ }\textbf {\bibinfo {volume} {56}},\ \bibinfo {pages} {25–30} (\bibinfo {year} {2003})}\BibitemShut {NoStop}%
\bibitem [{\citenamefont {Dagotto}(2005)}]{Dagotto2005}%
  \BibitemOpen
  \bibfield  {author} {\bibinfo {author} {\bibfnamefont {E.}~\bibnamefont {Dagotto}},\ }\href {https://www.science.org/doi/abs/10.1126/science.1107559} {\bibfield  {journal} {\bibinfo  {journal} {Science}\ }\textbf {\bibinfo {volume} {309}},\ \bibinfo {pages} {257} (\bibinfo {year} {2005})}\BibitemShut {NoStop}%
\bibitem [{\citenamefont {McElroy}\ \emph {et~al.}(2005)\citenamefont {McElroy}, \citenamefont {Lee}, \citenamefont {Slezak}, \citenamefont {Lee}, \citenamefont {Eisaki}, \citenamefont {Uchida},\ and\ \citenamefont {Davis}}]{McElroy2005}%
  \BibitemOpen
  \bibfield  {author} {\bibinfo {author} {\bibfnamefont {K.}~\bibnamefont {McElroy}}, \bibinfo {author} {\bibfnamefont {J.}~\bibnamefont {Lee}}, \bibinfo {author} {\bibfnamefont {J.~A.}\ \bibnamefont {Slezak}}, \bibinfo {author} {\bibfnamefont {D.-H.}\ \bibnamefont {Lee}}, \bibinfo {author} {\bibfnamefont {H.}~\bibnamefont {Eisaki}}, \bibinfo {author} {\bibfnamefont {S.}~\bibnamefont {Uchida}},\ and\ \bibinfo {author} {\bibfnamefont {J.~C.}\ \bibnamefont {Davis}},\ }\href {https://www.science.org/doi/abs/10.1126/science.1113095} {\bibfield  {journal} {\bibinfo  {journal} {Science}\ }\textbf {\bibinfo {volume} {309}},\ \bibinfo {pages} {1048} (\bibinfo {year} {2005})}\BibitemShut {NoStop}%
\bibitem [{\citenamefont {Qazilbash}\ \emph {et~al.}(2007)\citenamefont {Qazilbash}, \citenamefont {Brehm}, \citenamefont {Chae}, \citenamefont {Ho}, \citenamefont {Andreev}, \citenamefont {Kim}, \citenamefont {Yun}, \citenamefont {Balatsky}, \citenamefont {Maple}, \citenamefont {Keilmann}, \citenamefont {Kim},\ and\ \citenamefont {Basov}}]{Qazilbash2007}%
  \BibitemOpen
  \bibfield  {author} {\bibinfo {author} {\bibfnamefont {M.~M.}\ \bibnamefont {Qazilbash}}, \bibinfo {author} {\bibfnamefont {M.}~\bibnamefont {Brehm}}, \bibinfo {author} {\bibfnamefont {B.-G.}\ \bibnamefont {Chae}}, \bibinfo {author} {\bibfnamefont {P.-C.}\ \bibnamefont {Ho}}, \bibinfo {author} {\bibfnamefont {G.~O.}\ \bibnamefont {Andreev}}, \bibinfo {author} {\bibfnamefont {B.-J.}\ \bibnamefont {Kim}}, \bibinfo {author} {\bibfnamefont {S.~J.}\ \bibnamefont {Yun}}, \bibinfo {author} {\bibfnamefont {A.~V.}\ \bibnamefont {Balatsky}}, \bibinfo {author} {\bibfnamefont {M.~B.}\ \bibnamefont {Maple}}, \bibinfo {author} {\bibfnamefont {F.}~\bibnamefont {Keilmann}}, \bibinfo {author} {\bibfnamefont {H.-T.}\ \bibnamefont {Kim}},\ and\ \bibinfo {author} {\bibfnamefont {D.~N.}\ \bibnamefont {Basov}},\ }\href {https://doi.org/10.1126/science.1150124} {\bibfield  {journal} {\bibinfo  {journal} {Science}\ }\textbf {\bibinfo {volume} {318}},\ \bibinfo {pages} {1750–1753} (\bibinfo {year} {2007})}\BibitemShut {NoStop}%
\bibitem [{\citenamefont {Post}\ \emph {et~al.}(2018)\citenamefont {Post}, \citenamefont {McLeod}, \citenamefont {Hepting}, \citenamefont {Bluschke}, \citenamefont {Wang}, \citenamefont {Cristiani}, \citenamefont {Logvenov}, \citenamefont {Charnukha}, \citenamefont {Ni}, \citenamefont {Radhakrishnan}, \citenamefont {Minola}, \citenamefont {Pasupathy}, \citenamefont {Boris}, \citenamefont {Benckiser}, \citenamefont {Dahmen}, \citenamefont {Carlson}, \citenamefont {Keimer},\ and\ \citenamefont {Basov}}]{Post2018}%
  \BibitemOpen
  \bibfield  {author} {\bibinfo {author} {\bibfnamefont {K.~W.}\ \bibnamefont {Post}}, \bibinfo {author} {\bibfnamefont {A.~S.}\ \bibnamefont {McLeod}}, \bibinfo {author} {\bibfnamefont {M.}~\bibnamefont {Hepting}}, \bibinfo {author} {\bibfnamefont {M.}~\bibnamefont {Bluschke}}, \bibinfo {author} {\bibfnamefont {Y.}~\bibnamefont {Wang}}, \bibinfo {author} {\bibfnamefont {G.}~\bibnamefont {Cristiani}}, \bibinfo {author} {\bibfnamefont {G.}~\bibnamefont {Logvenov}}, \bibinfo {author} {\bibfnamefont {A.}~\bibnamefont {Charnukha}}, \bibinfo {author} {\bibfnamefont {G.~X.}\ \bibnamefont {Ni}}, \bibinfo {author} {\bibfnamefont {P.}~\bibnamefont {Radhakrishnan}}, \bibinfo {author} {\bibfnamefont {M.}~\bibnamefont {Minola}}, \bibinfo {author} {\bibfnamefont {A.}~\bibnamefont {Pasupathy}}, \bibinfo {author} {\bibfnamefont {A.~V.}\ \bibnamefont {Boris}}, \bibinfo {author} {\bibfnamefont {E.}~\bibnamefont {Benckiser}}, \bibinfo {author} {\bibfnamefont {K.~A.}\ \bibnamefont {Dahmen}}, \bibinfo {author} {\bibfnamefont
  {E.~W.}\ \bibnamefont {Carlson}}, \bibinfo {author} {\bibfnamefont {B.}~\bibnamefont {Keimer}},\ and\ \bibinfo {author} {\bibfnamefont {D.~N.}\ \bibnamefont {Basov}},\ }\href {https://doi.org/10.1038/s41567-018-0201-1} {\bibfield  {journal} {\bibinfo  {journal} {Nature Physics}\ }\textbf {\bibinfo {volume} {14}},\ \bibinfo {pages} {1056–1061} (\bibinfo {year} {2018})}\BibitemShut {NoStop}%
\bibitem [{\citenamefont {Shi}\ \emph {et~al.}(2020)\citenamefont {Shi}, \citenamefont {Xu}, \citenamefont {Yang}, \citenamefont {Slizovskiy}, \citenamefont {Morozov}, \citenamefont {Son}, \citenamefont {Ozdemir}, \citenamefont {Mullan}, \citenamefont {Barrier}, \citenamefont {Yin}, \citenamefont {Berdyugin}, \citenamefont {Piot}, \citenamefont {Taniguchi}, \citenamefont {Watanabe}, \citenamefont {Fal'ko}, \citenamefont {Novoselov}, \citenamefont {Geim},\ and\ \citenamefont {Mishchenko}}]{Shi2020}%
  \BibitemOpen
  \bibfield  {author} {\bibinfo {author} {\bibfnamefont {Y.}~\bibnamefont {Shi}}, \bibinfo {author} {\bibfnamefont {S.}~\bibnamefont {Xu}}, \bibinfo {author} {\bibfnamefont {Y.}~\bibnamefont {Yang}}, \bibinfo {author} {\bibfnamefont {S.}~\bibnamefont {Slizovskiy}}, \bibinfo {author} {\bibfnamefont {S.~V.}\ \bibnamefont {Morozov}}, \bibinfo {author} {\bibfnamefont {S.-K.}\ \bibnamefont {Son}}, \bibinfo {author} {\bibfnamefont {S.}~\bibnamefont {Ozdemir}}, \bibinfo {author} {\bibfnamefont {C.}~\bibnamefont {Mullan}}, \bibinfo {author} {\bibfnamefont {J.}~\bibnamefont {Barrier}}, \bibinfo {author} {\bibfnamefont {J.}~\bibnamefont {Yin}}, \bibinfo {author} {\bibfnamefont {A.~I.}\ \bibnamefont {Berdyugin}}, \bibinfo {author} {\bibfnamefont {B.~A.}\ \bibnamefont {Piot}}, \bibinfo {author} {\bibfnamefont {T.}~\bibnamefont {Taniguchi}}, \bibinfo {author} {\bibfnamefont {K.}~\bibnamefont {Watanabe}}, \bibinfo {author} {\bibfnamefont {V.~I.}\ \bibnamefont {Fal'ko}}, \bibinfo {author} {\bibfnamefont {K.~S.}\
  \bibnamefont {Novoselov}}, \bibinfo {author} {\bibfnamefont {A.~K.}\ \bibnamefont {Geim}},\ and\ \bibinfo {author} {\bibfnamefont {A.}~\bibnamefont {Mishchenko}},\ }\href {https://www.nature.com/articles/s41586-020-2568-2} {\bibfield  {journal} {\bibinfo  {journal} {Nature}\ }\textbf {\bibinfo {volume} {584}},\ \bibinfo {pages} {210} (\bibinfo {year} {2020})}\BibitemShut {NoStop}%
\bibitem [{\citenamefont {Song}\ \emph {et~al.}(2023)\citenamefont {Song}, \citenamefont {Main}, \citenamefont {Simmons}, \citenamefont {Liu}, \citenamefont {Phillabaum}, \citenamefont {Dahmen}, \citenamefont {Hudson}, \citenamefont {Hoffman},\ and\ \citenamefont {Carlson}}]{Song2023}%
  \BibitemOpen
  \bibfield  {author} {\bibinfo {author} {\bibfnamefont {C.-L.}\ \bibnamefont {Song}}, \bibinfo {author} {\bibfnamefont {E.~J.}\ \bibnamefont {Main}}, \bibinfo {author} {\bibfnamefont {F.}~\bibnamefont {Simmons}}, \bibinfo {author} {\bibfnamefont {S.}~\bibnamefont {Liu}}, \bibinfo {author} {\bibfnamefont {B.}~\bibnamefont {Phillabaum}}, \bibinfo {author} {\bibfnamefont {K.~A.}\ \bibnamefont {Dahmen}}, \bibinfo {author} {\bibfnamefont {E.~W.}\ \bibnamefont {Hudson}}, \bibinfo {author} {\bibfnamefont {J.~E.}\ \bibnamefont {Hoffman}},\ and\ \bibinfo {author} {\bibfnamefont {E.~W.}\ \bibnamefont {Carlson}},\ }\href {http://dx.doi.org/10.1038/s41467-023-38249-3} {\bibfield  {journal} {\bibinfo  {journal} {Nature Communications}\ }\textbf {\bibinfo {volume} {14}},\ \bibinfo {pages} {2622} (\bibinfo {year} {2023})}\BibitemShut {NoStop}%
\bibitem [{\citenamefont {Becker}\ \emph {et~al.}(2002)\citenamefont {Becker}, \citenamefont {Streng}, \citenamefont {Luo}, \citenamefont {Moshnyaga}, \citenamefont {Damaschke}, \citenamefont {Shannon},\ and\ \citenamefont {Samwer}}]{Becker2002}%
  \BibitemOpen
  \bibfield  {author} {\bibinfo {author} {\bibfnamefont {T.}~\bibnamefont {Becker}}, \bibinfo {author} {\bibfnamefont {C.}~\bibnamefont {Streng}}, \bibinfo {author} {\bibfnamefont {Y.}~\bibnamefont {Luo}}, \bibinfo {author} {\bibfnamefont {V.}~\bibnamefont {Moshnyaga}}, \bibinfo {author} {\bibfnamefont {B.}~\bibnamefont {Damaschke}}, \bibinfo {author} {\bibfnamefont {N.}~\bibnamefont {Shannon}},\ and\ \bibinfo {author} {\bibfnamefont {K.}~\bibnamefont {Samwer}},\ }\href {https://doi.org/10.1103/PhysRevLett.89.237203} {\bibfield  {journal} {\bibinfo  {journal} {Physical Review Letters}\ }\textbf {\bibinfo {volume} {89}},\ \bibinfo {pages} {237203} (\bibinfo {year} {2002})}\BibitemShut {NoStop}%
\bibitem [{\citenamefont {Dho}\ \emph {et~al.}(1999)\citenamefont {Dho}, \citenamefont {Kim}, \citenamefont {Lee}, \citenamefont {Kim}, \citenamefont {Lee}, \citenamefont {Jung},\ and\ \citenamefont {Noh}}]{Dho1999}%
  \BibitemOpen
  \bibfield  {author} {\bibinfo {author} {\bibfnamefont {J.}~\bibnamefont {Dho}}, \bibinfo {author} {\bibfnamefont {I.}~\bibnamefont {Kim}}, \bibinfo {author} {\bibfnamefont {S.}~\bibnamefont {Lee}}, \bibinfo {author} {\bibfnamefont {K.~H.}\ \bibnamefont {Kim}}, \bibinfo {author} {\bibfnamefont {H.~J.}\ \bibnamefont {Lee}}, \bibinfo {author} {\bibfnamefont {J.~H.}\ \bibnamefont {Jung}},\ and\ \bibinfo {author} {\bibfnamefont {T.~W.}\ \bibnamefont {Noh}},\ }\href {https://doi.org/10.1103/PhysRevB.59.492} {\bibfield  {journal} {\bibinfo  {journal} {Physical Review B}\ }\textbf {\bibinfo {volume} {59}},\ \bibinfo {pages} {492} (\bibinfo {year} {1999})}\BibitemShut {NoStop}%
\bibitem [{\citenamefont {Qazilbash}\ \emph {et~al.}(2011)\citenamefont {Qazilbash}, \citenamefont {Tripathi}, \citenamefont {Schafgans}, \citenamefont {Kim}, \citenamefont {Kim}, \citenamefont {Cai}, \citenamefont {Holt}, \citenamefont {Maser}, \citenamefont {Keilmann}, \citenamefont {Shpyrko},\ and\ \citenamefont {Basov}}]{Qazilbash2011}%
  \BibitemOpen
  \bibfield  {author} {\bibinfo {author} {\bibfnamefont {M.~M.}\ \bibnamefont {Qazilbash}}, \bibinfo {author} {\bibfnamefont {A.}~\bibnamefont {Tripathi}}, \bibinfo {author} {\bibfnamefont {A.~A.}\ \bibnamefont {Schafgans}}, \bibinfo {author} {\bibfnamefont {B.-J.}\ \bibnamefont {Kim}}, \bibinfo {author} {\bibfnamefont {H.-T.}\ \bibnamefont {Kim}}, \bibinfo {author} {\bibfnamefont {Z.}~\bibnamefont {Cai}}, \bibinfo {author} {\bibfnamefont {M.~V.}\ \bibnamefont {Holt}}, \bibinfo {author} {\bibfnamefont {J.~M.}\ \bibnamefont {Maser}}, \bibinfo {author} {\bibfnamefont {F.}~\bibnamefont {Keilmann}}, \bibinfo {author} {\bibfnamefont {O.~G.}\ \bibnamefont {Shpyrko}},\ and\ \bibinfo {author} {\bibfnamefont {D.~N.}\ \bibnamefont {Basov}},\ }\href {https://link.aps.org/doi/10.1103/PhysRevB.83.165108} {\bibfield  {journal} {\bibinfo  {journal} {Physical Review B}\ }\textbf {\bibinfo {volume} {83}},\ \bibinfo {pages} {165108} (\bibinfo {year} {2011})}\BibitemShut {NoStop}%
\bibitem [{\citenamefont {Kumar}\ \emph {et~al.}(2014)\citenamefont {Kumar}, \citenamefont {Strachan}, \citenamefont {Pickett}, \citenamefont {Bratkovsky}, \citenamefont {Nishi},\ and\ \citenamefont {Williams}}]{Kumar2014}%
  \BibitemOpen
  \bibfield  {author} {\bibinfo {author} {\bibfnamefont {S.}~\bibnamefont {Kumar}}, \bibinfo {author} {\bibfnamefont {J.~P.}\ \bibnamefont {Strachan}}, \bibinfo {author} {\bibfnamefont {M.~D.}\ \bibnamefont {Pickett}}, \bibinfo {author} {\bibfnamefont {A.}~\bibnamefont {Bratkovsky}}, \bibinfo {author} {\bibfnamefont {Y.}~\bibnamefont {Nishi}},\ and\ \bibinfo {author} {\bibfnamefont {R.~S.}\ \bibnamefont {Williams}},\ }\href {https://doi.org/10.1002/adma.201402404} {\bibfield  {journal} {\bibinfo  {journal} {Advanced Materials}\ }\textbf {\bibinfo {volume} {26}},\ \bibinfo {pages} {7505–7509} (\bibinfo {year} {2014})}\BibitemShut {NoStop}%
\bibitem [{Note1()}]{Note1}%
  \BibitemOpen
  \bibinfo {note} {See {\protect \em e.g.}, Fig.~4 of Ref.~\cite {Becker2002}, Fig. 2 of Ref.~\cite {Rocco2022}, Fig. 4(b) of Ref.~\cite {rozenberg-RN-nickelates}, Fig. 4 of Ref.~\cite {Wang2015}, Fig. 4 of Ref.~\cite {Guenon2013}, and Figs. 4 and 10 of Ref.~\cite {Lange2021}.}\BibitemShut {Stop}%
\bibitem [{\citenamefont {Vardi}\ \emph {et~al.}(2017)\citenamefont {Vardi}, \citenamefont {Anouchi}, \citenamefont {Yamin}, \citenamefont {Middey}, \citenamefont {Kareev}, \citenamefont {Chakhalian}, \citenamefont {Dubi},\ and\ \citenamefont {Sharoni}}]{ramp-reversal}%
  \BibitemOpen
  \bibfield  {author} {\bibinfo {author} {\bibfnamefont {N.}~\bibnamefont {Vardi}}, \bibinfo {author} {\bibfnamefont {E.}~\bibnamefont {Anouchi}}, \bibinfo {author} {\bibfnamefont {T.}~\bibnamefont {Yamin}}, \bibinfo {author} {\bibfnamefont {S.}~\bibnamefont {Middey}}, \bibinfo {author} {\bibfnamefont {M.}~\bibnamefont {Kareev}}, \bibinfo {author} {\bibfnamefont {J.}~\bibnamefont {Chakhalian}}, \bibinfo {author} {\bibfnamefont {Y.}~\bibnamefont {Dubi}},\ and\ \bibinfo {author} {\bibfnamefont {A.}~\bibnamefont {Sharoni}},\ }\href {https://doi.org/10.1002/adma.201605029} {\bibfield  {journal} {\bibinfo  {journal} {Advanced Materials}\ }\textbf {\bibinfo {volume} {29}},\ \bibinfo {pages} {1605029 } (\bibinfo {year} {2017})}\BibitemShut {NoStop}%
\bibitem [{\citenamefont {Guénon}\ \emph {et~al.}(2013)\citenamefont {Guénon}, \citenamefont {Scharinger}, \citenamefont {Wang}, \citenamefont {Ramírez}, \citenamefont {Koelle}, \citenamefont {Kleiner},\ and\ \citenamefont {Schuller}}]{Guenon2013}%
  \BibitemOpen
  \bibfield  {author} {\bibinfo {author} {\bibfnamefont {S.}~\bibnamefont {Guénon}}, \bibinfo {author} {\bibfnamefont {S.}~\bibnamefont {Scharinger}}, \bibinfo {author} {\bibfnamefont {S.}~\bibnamefont {Wang}}, \bibinfo {author} {\bibfnamefont {J.~G.}\ \bibnamefont {Ramírez}}, \bibinfo {author} {\bibfnamefont {D.}~\bibnamefont {Koelle}}, \bibinfo {author} {\bibfnamefont {R.}~\bibnamefont {Kleiner}},\ and\ \bibinfo {author} {\bibfnamefont {I.~K.}\ \bibnamefont {Schuller}},\ }\href {https://doi.org/10.1209/0295-5075/101/57003} {\bibfield  {journal} {\bibinfo  {journal} {EPL (Europhysics Letters)}\ }\textbf {\bibinfo {volume} {101}},\ \bibinfo {pages} {57003} (\bibinfo {year} {2013})}\BibitemShut {NoStop}%
\bibitem [{\citenamefont {Sharoni}\ \emph {et~al.}(2008)\citenamefont {Sharoni}, \citenamefont {Ramírez},\ and\ \citenamefont {Schuller}}]{sharoni-avalanches}%
  \BibitemOpen
  \bibfield  {author} {\bibinfo {author} {\bibfnamefont {A.}~\bibnamefont {Sharoni}}, \bibinfo {author} {\bibfnamefont {J.~G.}\ \bibnamefont {Ramírez}},\ and\ \bibinfo {author} {\bibfnamefont {I.~K.}\ \bibnamefont {Schuller}},\ }\href {https://doi.org/10.1103/physrevlett.101.026404} {\bibfield  {journal} {\bibinfo  {journal} {Physical Review Letters}\ }\textbf {\bibinfo {volume} {101}},\ \bibinfo {pages} {026404} (\bibinfo {year} {2008})}\BibitemShut {NoStop}%
\bibitem [{\citenamefont {Liu}\ \emph {et~al.}(2016)\citenamefont {Liu}, \citenamefont {Phillabaum}, \citenamefont {Carlson}, \citenamefont {Dahmen}, \citenamefont {Vidhyadhiraja}, \citenamefont {Qazilbash},\ and\ \citenamefont {Basov}}]{shuo-prl}%
  \BibitemOpen
  \bibfield  {author} {\bibinfo {author} {\bibfnamefont {S.}~\bibnamefont {Liu}}, \bibinfo {author} {\bibfnamefont {B.}~\bibnamefont {Phillabaum}}, \bibinfo {author} {\bibfnamefont {E.~W.}\ \bibnamefont {Carlson}}, \bibinfo {author} {\bibfnamefont {K.~A.}\ \bibnamefont {Dahmen}}, \bibinfo {author} {\bibfnamefont {N.~S.}\ \bibnamefont {Vidhyadhiraja}}, \bibinfo {author} {\bibfnamefont {M.~M.}\ \bibnamefont {Qazilbash}},\ and\ \bibinfo {author} {\bibfnamefont {D.~N.}\ \bibnamefont {Basov}},\ }\href {https://doi.org/10.1103/physrevlett.116.036401} {\bibfield  {journal} {\bibinfo  {journal} {Physical Review Letters}\ }\textbf {\bibinfo {volume} {116}},\ \bibinfo {pages} {036401 } (\bibinfo {year} {2016})}\BibitemShut {NoStop}%
\bibitem [{\citenamefont {Basak}\ \emph {et~al.}(2023{\natexlab{a}})\citenamefont {Basak}, \citenamefont {Banguero}, \citenamefont {Burzawa}, \citenamefont {Simmons}, \citenamefont {Salev}, \citenamefont {Aigouy}, \citenamefont {Qazilbash}, \citenamefont {Schuller}, \citenamefont {Basov}, \citenamefont {Zimmers},\ and\ \citenamefont {Carlson}}]{basak-deep-learning}%
  \BibitemOpen
  \bibfield  {author} {\bibinfo {author} {\bibfnamefont {S.}~\bibnamefont {Basak}}, \bibinfo {author} {\bibfnamefont {M.~A.}\ \bibnamefont {Banguero}}, \bibinfo {author} {\bibfnamefont {L.}~\bibnamefont {Burzawa}}, \bibinfo {author} {\bibfnamefont {F.}~\bibnamefont {Simmons}}, \bibinfo {author} {\bibfnamefont {P.}~\bibnamefont {Salev}}, \bibinfo {author} {\bibfnamefont {L.}~\bibnamefont {Aigouy}}, \bibinfo {author} {\bibfnamefont {M.~M.}\ \bibnamefont {Qazilbash}}, \bibinfo {author} {\bibfnamefont {I.~K.}\ \bibnamefont {Schuller}}, \bibinfo {author} {\bibfnamefont {D.~N.}\ \bibnamefont {Basov}}, \bibinfo {author} {\bibfnamefont {A.}~\bibnamefont {Zimmers}},\ and\ \bibinfo {author} {\bibfnamefont {E.~W.}\ \bibnamefont {Carlson}},\ }\href {https://doi.org/10.1103/physrevb.107.205121} {\bibfield  {journal} {\bibinfo  {journal} {Physical Review B}\ }\textbf {\bibinfo {volume} {107}},\ \bibinfo {pages} {205121} (\bibinfo {year} {2023}{\natexlab{a}})}\BibitemShut {NoStop}%
\bibitem [{\citenamefont {Alzate~Banguero}\ \emph {et~al.}(2025)\citenamefont {Alzate~Banguero}, \citenamefont {Basak}, \citenamefont {Raymond}, \citenamefont {Simmons}, \citenamefont {Salev}, \citenamefont {Schuller}, \citenamefont {Aigouy}, \citenamefont {Carlson},\ and\ \citenamefont {Zimmers}}]{AlzateBanguero2025}%
  \BibitemOpen
  \bibfield  {author} {\bibinfo {author} {\bibfnamefont {M.}~\bibnamefont {Alzate~Banguero}}, \bibinfo {author} {\bibfnamefont {S.}~\bibnamefont {Basak}}, \bibinfo {author} {\bibfnamefont {N.}~\bibnamefont {Raymond}}, \bibinfo {author} {\bibfnamefont {F.}~\bibnamefont {Simmons}}, \bibinfo {author} {\bibfnamefont {P.}~\bibnamefont {Salev}}, \bibinfo {author} {\bibfnamefont {I.~K.}\ \bibnamefont {Schuller}}, \bibinfo {author} {\bibfnamefont {L.}~\bibnamefont {Aigouy}}, \bibinfo {author} {\bibfnamefont {E.~W.}\ \bibnamefont {Carlson}},\ and\ \bibinfo {author} {\bibfnamefont {A.}~\bibnamefont {Zimmers}},\ }\href {https://doi.org/10.3390/condmat10010012} {\bibfield  {journal} {\bibinfo  {journal} {Condensed Matter}\ }\textbf {\bibinfo {volume} {10}},\ \bibinfo {pages} {12} (\bibinfo {year} {2025})}\BibitemShut {NoStop}%
\bibitem [{\citenamefont {Kepič}\ \emph {et~al.}(2025)\citenamefont {Kepič}, \citenamefont {Horák}, \citenamefont {Kabát}, \citenamefont {Hájek}, \citenamefont {Konečná}, \citenamefont {Šikola},\ and\ \citenamefont {Ligmajer}}]{Kepi2025}%
  \BibitemOpen
  \bibfield  {author} {\bibinfo {author} {\bibfnamefont {P.}~\bibnamefont {Kepič}}, \bibinfo {author} {\bibfnamefont {M.}~\bibnamefont {Horák}}, \bibinfo {author} {\bibfnamefont {J.}~\bibnamefont {Kabát}}, \bibinfo {author} {\bibfnamefont {M.}~\bibnamefont {Hájek}}, \bibinfo {author} {\bibfnamefont {A.}~\bibnamefont {Konečná}}, \bibinfo {author} {\bibfnamefont {T.}~\bibnamefont {Šikola}},\ and\ \bibinfo {author} {\bibfnamefont {F.}~\bibnamefont {Ligmajer}},\ }\href {https://doi.org/10.1021/acsnano.4c13188} {\bibfield  {journal} {\bibinfo  {journal} {ACS Nano}\ }\textbf {\bibinfo {volume} {19}},\ \bibinfo {pages} {1167–1176} (\bibinfo {year} {2025})}\BibitemShut {NoStop}%
\bibitem [{\citenamefont {Liu}\ \emph {et~al.}(2013)\citenamefont {Liu}, \citenamefont {Wagner}, \citenamefont {Abreu}, \citenamefont {Kittiwatanakul}, \citenamefont {McLeod}, \citenamefont {Fei}, \citenamefont {Goldflam}, \citenamefont {Dai}, \citenamefont {Fogler}, \citenamefont {Lu},\ and\ \citenamefont {et~al.}}]{Liu2013}%
  \BibitemOpen
  \bibfield  {author} {\bibinfo {author} {\bibfnamefont {M.~K.}\ \bibnamefont {Liu}}, \bibinfo {author} {\bibfnamefont {M.}~\bibnamefont {Wagner}}, \bibinfo {author} {\bibfnamefont {E.}~\bibnamefont {Abreu}}, \bibinfo {author} {\bibfnamefont {S.}~\bibnamefont {Kittiwatanakul}}, \bibinfo {author} {\bibfnamefont {A.}~\bibnamefont {McLeod}}, \bibinfo {author} {\bibfnamefont {Z.}~\bibnamefont {Fei}}, \bibinfo {author} {\bibfnamefont {M.}~\bibnamefont {Goldflam}}, \bibinfo {author} {\bibfnamefont {S.}~\bibnamefont {Dai}}, \bibinfo {author} {\bibfnamefont {M.~M.}\ \bibnamefont {Fogler}}, \bibinfo {author} {\bibfnamefont {J.}~\bibnamefont {Lu}},\ and\ \bibinfo {author} {\bibnamefont {et~al.}},\ }\href {https://doi.org/10.1103/PhysRevLett.111.096602} {\bibfield  {journal} {\bibinfo  {journal} {Phys. Rev. Lett.}\ }\textbf {\bibinfo {volume} {111}},\ \bibinfo {pages} {096602} (\bibinfo {year} {2013})}\BibitemShut {NoStop}%
\bibitem [{\citenamefont {Stinson}\ \emph {et~al.}(2018)\citenamefont {Stinson}, \citenamefont {Sternbach}, \citenamefont {Najera}, \citenamefont {Jing}, \citenamefont {McLeod}, \citenamefont {Slusar}, \citenamefont {Mueller}, \citenamefont {Anderegg}, \citenamefont {Kim}, \citenamefont {Rozenberg},\ and\ \citenamefont {et~al.}}]{Stinson2018}%
  \BibitemOpen
  \bibfield  {author} {\bibinfo {author} {\bibfnamefont {H.~T.}\ \bibnamefont {Stinson}}, \bibinfo {author} {\bibfnamefont {A.}~\bibnamefont {Sternbach}}, \bibinfo {author} {\bibfnamefont {O.}~\bibnamefont {Najera}}, \bibinfo {author} {\bibfnamefont {R.}~\bibnamefont {Jing}}, \bibinfo {author} {\bibfnamefont {A.~S.}\ \bibnamefont {McLeod}}, \bibinfo {author} {\bibfnamefont {T.~V.}\ \bibnamefont {Slusar}}, \bibinfo {author} {\bibfnamefont {A.}~\bibnamefont {Mueller}}, \bibinfo {author} {\bibfnamefont {L.}~\bibnamefont {Anderegg}}, \bibinfo {author} {\bibfnamefont {H.-T.}\ \bibnamefont {Kim}}, \bibinfo {author} {\bibfnamefont {M.}~\bibnamefont {Rozenberg}},\ and\ \bibinfo {author} {\bibnamefont {et~al.}},\ }\href {https://doi.org/10.1038/s41467-018-05998-5} {\bibfield  {journal} {\bibinfo  {journal} {Nat. Commun.}\ }\textbf {\bibinfo {volume} {9}},\ \bibinfo {pages} {4598} (\bibinfo {year} {2018})}\BibitemShut {NoStop}%
\bibitem [{\citenamefont {Spitzig}\ \emph {et~al.}(2022)\citenamefont {Spitzig}, \citenamefont {Pivonka}, \citenamefont {Frenzel}, \citenamefont {Kim}, \citenamefont {Ko}, \citenamefont {Zhou}, \citenamefont {Hudson}, \citenamefont {Ramanathan}, \citenamefont {Hoffman},\ and\ \citenamefont {Hoffman}}]{Spitzig2022}%
  \BibitemOpen
  \bibfield  {author} {\bibinfo {author} {\bibfnamefont {A.}~\bibnamefont {Spitzig}}, \bibinfo {author} {\bibfnamefont {A.}~\bibnamefont {Pivonka}}, \bibinfo {author} {\bibfnamefont {A.}~\bibnamefont {Frenzel}}, \bibinfo {author} {\bibfnamefont {J.}~\bibnamefont {Kim}}, \bibinfo {author} {\bibfnamefont {C.}~\bibnamefont {Ko}}, \bibinfo {author} {\bibfnamefont {Y.}~\bibnamefont {Zhou}}, \bibinfo {author} {\bibfnamefont {E.}~\bibnamefont {Hudson}}, \bibinfo {author} {\bibfnamefont {S.}~\bibnamefont {Ramanathan}}, \bibinfo {author} {\bibfnamefont {J.~E.}\ \bibnamefont {Hoffman}},\ and\ \bibinfo {author} {\bibfnamefont {J.~D.}\ \bibnamefont {Hoffman}},\ }\href {https://doi.org/10.1063/5.0086932} {\bibfield  {journal} {\bibinfo  {journal} {Appl. Phys. Lett.}\ }\textbf {\bibinfo {volume} {120}},\ \bibinfo {pages} {151602} (\bibinfo {year} {2022})}\BibitemShut {NoStop}%
\bibitem [{\citenamefont {Wang}\ \emph {et~al.}(2015)\citenamefont {Wang}, \citenamefont {Ramírez},\ and\ \citenamefont {Schuller}}]{Wang2015}%
  \BibitemOpen
  \bibfield  {author} {\bibinfo {author} {\bibfnamefont {S.}~\bibnamefont {Wang}}, \bibinfo {author} {\bibfnamefont {J.~G.}\ \bibnamefont {Ramírez}},\ and\ \bibinfo {author} {\bibfnamefont {I.~K.}\ \bibnamefont {Schuller}},\ }\href {http://dx.doi.org/10.1103/PhysRevB.92.085150} {\bibfield  {journal} {\bibinfo  {journal} {Physical Review B}\ }\textbf {\bibinfo {volume} {92}},\ \bibinfo {pages} {085150} (\bibinfo {year} {2015})}\BibitemShut {NoStop}%
\bibitem [{\citenamefont {Rocco}\ \emph {et~al.}(2022)\citenamefont {Rocco}, \citenamefont {del Valle}, \citenamefont {Navarro}, \citenamefont {Salev}, \citenamefont {Schuller},\ and\ \citenamefont {Rozenberg}}]{Rocco2022}%
  \BibitemOpen
  \bibfield  {author} {\bibinfo {author} {\bibfnamefont {R.}~\bibnamefont {Rocco}}, \bibinfo {author} {\bibfnamefont {J.}~\bibnamefont {del Valle}}, \bibinfo {author} {\bibfnamefont {H.}~\bibnamefont {Navarro}}, \bibinfo {author} {\bibfnamefont {P.}~\bibnamefont {Salev}}, \bibinfo {author} {\bibfnamefont {I.~K.}\ \bibnamefont {Schuller}},\ and\ \bibinfo {author} {\bibfnamefont {M.}~\bibnamefont {Rozenberg}},\ }\href {https://doi.org/10.1103/PhysRevApplied.17.024028} {\bibfield  {journal} {\bibinfo  {journal} {Phys. Rev. Appl.}\ }\textbf {\bibinfo {volume} {17}},\ \bibinfo {pages} {024028} (\bibinfo {year} {2022})}\BibitemShut {NoStop}%
\bibitem [{\citenamefont {Stoliar}\ \emph {et~al.}(2013)\citenamefont {Stoliar}, \citenamefont {Cario}, \citenamefont {Janod}, \citenamefont {Corraze}, \citenamefont {Guillot‐Deudon}, \citenamefont {Salmon‐Bourmand}, \citenamefont {Guiot}, \citenamefont {Tranchant},\ and\ \citenamefont {Rozenberg}}]{Stoliar2013}%
  \BibitemOpen
  \bibfield  {author} {\bibinfo {author} {\bibfnamefont {P.}~\bibnamefont {Stoliar}}, \bibinfo {author} {\bibfnamefont {L.}~\bibnamefont {Cario}}, \bibinfo {author} {\bibfnamefont {E.}~\bibnamefont {Janod}}, \bibinfo {author} {\bibfnamefont {B.}~\bibnamefont {Corraze}}, \bibinfo {author} {\bibfnamefont {C.}~\bibnamefont {Guillot‐Deudon}}, \bibinfo {author} {\bibfnamefont {S.}~\bibnamefont {Salmon‐Bourmand}}, \bibinfo {author} {\bibfnamefont {V.}~\bibnamefont {Guiot}}, \bibinfo {author} {\bibfnamefont {J.}~\bibnamefont {Tranchant}},\ and\ \bibinfo {author} {\bibfnamefont {M.}~\bibnamefont {Rozenberg}},\ }\href {https://doi.org/10.1002/adma.201301113} {\bibfield  {journal} {\bibinfo  {journal} {Advanced Materials}\ }\textbf {\bibinfo {volume} {25}},\ \bibinfo {pages} {3222} (\bibinfo {year} {2013})}\BibitemShut {NoStop}%
\bibitem [{\citenamefont {Lange}\ \emph {et~al.}(2021)\citenamefont {Lange}, \citenamefont {Guénon}, \citenamefont {Kalcheim}, \citenamefont {Luibrand}, \citenamefont {Vargas}, \citenamefont {Schwebius}, \citenamefont {Kleiner}, \citenamefont {Schuller},\ and\ \citenamefont {Koelle}}]{Lange2021}%
  \BibitemOpen
  \bibfield  {author} {\bibinfo {author} {\bibfnamefont {M.}~\bibnamefont {Lange}}, \bibinfo {author} {\bibfnamefont {S.}~\bibnamefont {Guénon}}, \bibinfo {author} {\bibfnamefont {Y.}~\bibnamefont {Kalcheim}}, \bibinfo {author} {\bibfnamefont {T.}~\bibnamefont {Luibrand}}, \bibinfo {author} {\bibfnamefont {N.~M.}\ \bibnamefont {Vargas}}, \bibinfo {author} {\bibfnamefont {D.}~\bibnamefont {Schwebius}}, \bibinfo {author} {\bibfnamefont {R.}~\bibnamefont {Kleiner}}, \bibinfo {author} {\bibfnamefont {I.~K.}\ \bibnamefont {Schuller}},\ and\ \bibinfo {author} {\bibfnamefont {D.}~\bibnamefont {Koelle}},\ }\href {http://dx.doi.org/10.1103/PhysRevApplied.16.054027} {\bibfield  {journal} {\bibinfo  {journal} {Physical Review Applied}\ }\textbf {\bibinfo {volume} {16}},\ \bibinfo {pages} {054027} (\bibinfo {year} {2021})}\BibitemShut {NoStop}%
\bibitem [{\citenamefont {Preisach}(1935)}]{Preisach1935}%
  \BibitemOpen
  \bibfield  {author} {\bibinfo {author} {\bibfnamefont {F.}~\bibnamefont {Preisach}},\ }\href {https://doi.org/10.1007/BF01349418} {\bibfield  {journal} {\bibinfo  {journal} {Zeitschrift für Physik}\ }\textbf {\bibinfo {volume} {94}},\ \bibinfo {pages} {277} (\bibinfo {year} {1935})}\BibitemShut {NoStop}%
\bibitem [{\citenamefont {Ram\'{\i}rez}\ \emph {et~al.}(2009)\citenamefont {Ram\'{\i}rez}, \citenamefont {Sharoni}, \citenamefont {Dubi}, \citenamefont {G\'omez},\ and\ \citenamefont {Schuller}}]{PhysRevB.79.235110}%
  \BibitemOpen
  \bibfield  {author} {\bibinfo {author} {\bibfnamefont {J.-G.}\ \bibnamefont {Ram\'{\i}rez}}, \bibinfo {author} {\bibfnamefont {A.}~\bibnamefont {Sharoni}}, \bibinfo {author} {\bibfnamefont {Y.}~\bibnamefont {Dubi}}, \bibinfo {author} {\bibfnamefont {M.~E.}\ \bibnamefont {G\'omez}},\ and\ \bibinfo {author} {\bibfnamefont {I.~K.}\ \bibnamefont {Schuller}},\ }\href {https://doi.org/10.1103/PhysRevB.79.235110} {\bibfield  {journal} {\bibinfo  {journal} {Phys. Rev. B}\ }\textbf {\bibinfo {volume} {79}},\ \bibinfo {pages} {235110} (\bibinfo {year} {2009})}\BibitemShut {NoStop}%
\bibitem [{\citenamefont {Suárez-Villagrán}\ \emph {et~al.}(2020)\citenamefont {Suárez-Villagrán}, \citenamefont {Mitsakos}, \citenamefont {Lee}, \citenamefont {Dobrosavljević}, \citenamefont {Miller},\ and\ \citenamefont {Miranda}}]{vlad-RF-RN}%
  \BibitemOpen
  \bibfield  {author} {\bibinfo {author} {\bibfnamefont {M.~Y.}\ \bibnamefont {Suárez-Villagrán}}, \bibinfo {author} {\bibfnamefont {N.}~\bibnamefont {Mitsakos}}, \bibinfo {author} {\bibfnamefont {T.-H.}\ \bibnamefont {Lee}}, \bibinfo {author} {\bibfnamefont {V.}~\bibnamefont {Dobrosavljević}}, \bibinfo {author} {\bibfnamefont {J.~H.}\ \bibnamefont {Miller}},\ and\ \bibinfo {author} {\bibfnamefont {E.}~\bibnamefont {Miranda}},\ }\href {https://doi.org/10.1103/physrevb.101.235112} {\bibfield  {journal} {\bibinfo  {journal} {Physical Review B}\ }\textbf {\bibinfo {volume} {101}},\ \bibinfo {pages} {235112} (\bibinfo {year} {2020})}\BibitemShut {NoStop}%
\bibitem [{\citenamefont {Frank}\ and\ \citenamefont {Lobb}(1988)}]{bp-algo}%
  \BibitemOpen
  \bibfield  {author} {\bibinfo {author} {\bibfnamefont {D.~J.}\ \bibnamefont {Frank}}\ and\ \bibinfo {author} {\bibfnamefont {C.~J.}\ \bibnamefont {Lobb}},\ }\href {https://doi.org/10.1103/PhysRevB.37.302} {\bibfield  {journal} {\bibinfo  {journal} {Phys. Rev. B}\ }\textbf {\bibinfo {volume} {37}},\ \bibinfo {pages} {302} (\bibinfo {year} {1988})}\BibitemShut {NoStop}%
\bibitem [{\citenamefont {Basak}\ \emph {et~al.}(2023{\natexlab{b}})\citenamefont {Basak}, \citenamefont {Sun}, \citenamefont {Banguero}, \citenamefont {Salev}, \citenamefont {Schuller}, \citenamefont {Aigouy}, \citenamefont {Carlson},\ and\ \citenamefont {Zimmers}}]{Basak2023}%
  \BibitemOpen
  \bibfield  {author} {\bibinfo {author} {\bibfnamefont {S.}~\bibnamefont {Basak}}, \bibinfo {author} {\bibfnamefont {Y.}~\bibnamefont {Sun}}, \bibinfo {author} {\bibfnamefont {M.~A.}\ \bibnamefont {Banguero}}, \bibinfo {author} {\bibfnamefont {P.}~\bibnamefont {Salev}}, \bibinfo {author} {\bibfnamefont {I.~K.}\ \bibnamefont {Schuller}}, \bibinfo {author} {\bibfnamefont {L.}~\bibnamefont {Aigouy}}, \bibinfo {author} {\bibfnamefont {E.~W.}\ \bibnamefont {Carlson}},\ and\ \bibinfo {author} {\bibfnamefont {A.}~\bibnamefont {Zimmers}},\ }\href {http://dx.doi.org/10.1002/aelm.202300085} {\bibfield  {journal} {\bibinfo  {journal} {Advanced Electronic Materials}\ }\textbf {\bibinfo {volume} {9}},\ \bibinfo {pages} {2300085} (\bibinfo {year} {2023}{\natexlab{b}})}\BibitemShut {NoStop}%
\bibitem [{Note2()}]{Note2}%
  \BibitemOpen
  \bibinfo {note} {For comparison, Guénon {\protect \em et al.} use a resistor network of size $20\times 400$ nodes, each with 4 resistors.\cite {Guenon2013})}\BibitemShut {NoStop}%
\bibitem [{\citenamefont {Ahmadi}\ \emph {et~al.}(2024)\citenamefont {Ahmadi}, \citenamefont {Atul}, \citenamefont {de~Graaf}, \citenamefont {van~der Veer}, \citenamefont {Meise}, \citenamefont {Tavabi}, \citenamefont {Heggen}, \citenamefont {Dunin-Borkowski}, \citenamefont {Ahmadi},\ and\ \citenamefont {Kooi}}]{Ahmadi2024}%
  \BibitemOpen
  \bibfield  {author} {\bibinfo {author} {\bibfnamefont {M.}~\bibnamefont {Ahmadi}}, \bibinfo {author} {\bibfnamefont {A.}~\bibnamefont {Atul}}, \bibinfo {author} {\bibfnamefont {S.}~\bibnamefont {de~Graaf}}, \bibinfo {author} {\bibfnamefont {E.}~\bibnamefont {van~der Veer}}, \bibinfo {author} {\bibfnamefont {A.}~\bibnamefont {Meise}}, \bibinfo {author} {\bibfnamefont {A.~H.}\ \bibnamefont {Tavabi}}, \bibinfo {author} {\bibfnamefont {M.}~\bibnamefont {Heggen}}, \bibinfo {author} {\bibfnamefont {R.~E.}\ \bibnamefont {Dunin-Borkowski}}, \bibinfo {author} {\bibfnamefont {M.}~\bibnamefont {Ahmadi}},\ and\ \bibinfo {author} {\bibfnamefont {B.~J.}\ \bibnamefont {Kooi}},\ }\href {https://doi.org/10.1021/acsnano.3c10745} {\bibfield  {journal} {\bibinfo  {journal} {ACS Nano}\ }\textbf {\bibinfo {volume} {18}},\ \bibinfo {pages} {13496–13505} (\bibinfo {year} {2024})}\BibitemShut {NoStop}%
\bibitem [{\citenamefont {Jian}\ \emph {et~al.}(2015)\citenamefont {Jian}, \citenamefont {Zhang}, \citenamefont {Jacob}, \citenamefont {Chen}, \citenamefont {Wang}, \citenamefont {Huang},\ and\ \citenamefont {Wang}}]{Jian2015}%
  \BibitemOpen
  \bibfield  {author} {\bibinfo {author} {\bibfnamefont {J.}~\bibnamefont {Jian}}, \bibinfo {author} {\bibfnamefont {W.}~\bibnamefont {Zhang}}, \bibinfo {author} {\bibfnamefont {C.}~\bibnamefont {Jacob}}, \bibinfo {author} {\bibfnamefont {A.}~\bibnamefont {Chen}}, \bibinfo {author} {\bibfnamefont {H.}~\bibnamefont {Wang}}, \bibinfo {author} {\bibfnamefont {J.}~\bibnamefont {Huang}},\ and\ \bibinfo {author} {\bibfnamefont {H.}~\bibnamefont {Wang}},\ }\href {http://dx.doi.org/10.1063/1.4930831} {\bibfield  {journal} {\bibinfo  {journal} {Applied Physics Letters}\ }\textbf {\bibinfo {volume} {107}},\ \bibinfo {pages} {102105} (\bibinfo {year} {2015})}\BibitemShut {NoStop}%
\bibitem [{\citenamefont {Li}\ \emph {et~al.}(2019)\citenamefont {Li}, \citenamefont {Pelliciari}, \citenamefont {Mazzoli}, \citenamefont {Catalano}, \citenamefont {Simmons}, \citenamefont {Sadowski}, \citenamefont {Levitan}, \citenamefont {Gibert}, \citenamefont {Carlson}, \citenamefont {Triscone}, \citenamefont {Wilkins}, \citenamefont {Comin}, \citenamefont {Gibert}, \citenamefont {Carlson}, \citenamefont {Triscone}, \citenamefont {Wilkins},\ and\ \citenamefont {Comin}}]{comin-ndnio3}%
  \BibitemOpen
  \bibfield  {author} {\bibinfo {author} {\bibfnamefont {J.}~\bibnamefont {Li}}, \bibinfo {author} {\bibfnamefont {J.}~\bibnamefont {Pelliciari}}, \bibinfo {author} {\bibfnamefont {C.}~\bibnamefont {Mazzoli}}, \bibinfo {author} {\bibfnamefont {S.}~\bibnamefont {Catalano}}, \bibinfo {author} {\bibfnamefont {F.}~\bibnamefont {Simmons}}, \bibinfo {author} {\bibfnamefont {J.~T.}\ \bibnamefont {Sadowski}}, \bibinfo {author} {\bibfnamefont {A.}~\bibnamefont {Levitan}}, \bibinfo {author} {\bibfnamefont {M.}~\bibnamefont {Gibert}}, \bibinfo {author} {\bibfnamefont {E.}~\bibnamefont {Carlson}}, \bibinfo {author} {\bibfnamefont {J.-M.}\ \bibnamefont {Triscone}}, \bibinfo {author} {\bibfnamefont {S.}~\bibnamefont {Wilkins}}, \bibinfo {author} {\bibfnamefont {R.}~\bibnamefont {Comin}}, \bibinfo {author} {\bibfnamefont {M.}~\bibnamefont {Gibert}}, \bibinfo {author} {\bibfnamefont {E.~W.}\ \bibnamefont {Carlson}}, \bibinfo {author} {\bibfnamefont {J.-M.}\ \bibnamefont {Triscone}}, \bibinfo {author} {\bibfnamefont
  {S.}~\bibnamefont {Wilkins}},\ and\ \bibinfo {author} {\bibfnamefont {R.}~\bibnamefont {Comin}},\ }\href {https://doi.org/10.1038/s41467-019-12502-0} {\bibfield  {journal} {\bibinfo  {journal} {Nature Communications}\ }\textbf {\bibinfo {volume} {10}},\ \bibinfo {pages} {4568} (\bibinfo {year} {2019})}\BibitemShut {NoStop}%
\bibitem [{\citenamefont {Bear}\ and\ \citenamefont {Bachmat}(1990)}]{porous}%
  \BibitemOpen
  \bibfield  {author} {\bibinfo {author} {\bibfnamefont {J.}~\bibnamefont {Bear}}\ and\ \bibinfo {author} {\bibfnamefont {Y.}~\bibnamefont {Bachmat}},\ }\href {https://link.springer.com/book/10.1007/978-94-009-1926-6} {\emph {\bibinfo {title} {Introduction to modeling of transport phenomena in porous media}}}\ (\bibinfo  {publisher} {Springer Dordrecht},\ \bibinfo {address} {The Netherlands},\ \bibinfo {year} {1990})\BibitemShut {NoStop}%
\bibitem [{\citenamefont {Hacker}\ \emph {et~al.}(2014)\citenamefont {Hacker}, \citenamefont {Yang},\ and\ \citenamefont {McCartney}}]{rcac-purdue}%
  \BibitemOpen
  \bibfield  {author} {\bibinfo {author} {\bibfnamefont {T.}~\bibnamefont {Hacker}}, \bibinfo {author} {\bibfnamefont {B.}~\bibnamefont {Yang}},\ and\ \bibinfo {author} {\bibfnamefont {G.}~\bibnamefont {McCartney}},\ }\href {https://er.educause.edu/articles/2014/7/empowering-faculty-a-campus-cyberinfrastructure-strategy-for-research-communities} {\bibfield  {journal} {\bibinfo  {journal} {Educause Review}\ } (\bibinfo {year} {2014})}\BibitemShut {NoStop}%
\bibitem [{\citenamefont {Luibrand}\ \emph {et~al.}(2023)\citenamefont {Luibrand}, \citenamefont {Bercher}, \citenamefont {Rocco}, \citenamefont {Tahouni-Bonab}, \citenamefont {Varbaro}, \citenamefont {Rischau}, \citenamefont {Domínguez}, \citenamefont {Zhou}, \citenamefont {Luo}, \citenamefont {Bag}, \citenamefont {Fratino}, \citenamefont {Kleiner}, \citenamefont {Gariglio}, \citenamefont {Koelle}, \citenamefont {Triscone}, \citenamefont {Rozenberg}, \citenamefont {Kuzmenko}, \citenamefont {Guénon},\ and\ \citenamefont {Valle}}]{rozenberg-RN-nickelates}%
  \BibitemOpen
  \bibfield  {author} {\bibinfo {author} {\bibfnamefont {T.}~\bibnamefont {Luibrand}}, \bibinfo {author} {\bibfnamefont {A.}~\bibnamefont {Bercher}}, \bibinfo {author} {\bibfnamefont {R.}~\bibnamefont {Rocco}}, \bibinfo {author} {\bibfnamefont {F.}~\bibnamefont {Tahouni-Bonab}}, \bibinfo {author} {\bibfnamefont {L.}~\bibnamefont {Varbaro}}, \bibinfo {author} {\bibfnamefont {C.~W.}\ \bibnamefont {Rischau}}, \bibinfo {author} {\bibfnamefont {C.}~\bibnamefont {Domínguez}}, \bibinfo {author} {\bibfnamefont {Y.}~\bibnamefont {Zhou}}, \bibinfo {author} {\bibfnamefont {W.}~\bibnamefont {Luo}}, \bibinfo {author} {\bibfnamefont {S.}~\bibnamefont {Bag}}, \bibinfo {author} {\bibfnamefont {L.}~\bibnamefont {Fratino}}, \bibinfo {author} {\bibfnamefont {R.}~\bibnamefont {Kleiner}}, \bibinfo {author} {\bibfnamefont {S.}~\bibnamefont {Gariglio}}, \bibinfo {author} {\bibfnamefont {D.}~\bibnamefont {Koelle}}, \bibinfo {author} {\bibfnamefont {J.-M.}\ \bibnamefont {Triscone}}, \bibinfo {author} {\bibfnamefont {M.~J.}\
  \bibnamefont {Rozenberg}}, \bibinfo {author} {\bibfnamefont {A.~B.}\ \bibnamefont {Kuzmenko}}, \bibinfo {author} {\bibfnamefont {S.}~\bibnamefont {Guénon}},\ and\ \bibinfo {author} {\bibfnamefont {J.~d.}\ \bibnamefont {Valle}},\ }\href {https://doi.org/10.1103/physrevresearch.5.013108} {\bibfield  {journal} {\bibinfo  {journal} {Physical Review Research}\ }\textbf {\bibinfo {volume} {5}},\ \bibinfo {pages} {013108} (\bibinfo {year} {2023})}\BibitemShut {NoStop}%
\end{thebibliography}
%

\clearpage

\newpage

\section{\label{sec:si} Supplementary Information}

\onecolumngrid

\section{Experimental data}

\begin{figure}[htbp]
    \centering
    \includegraphics[width=0.6\textwidth]{plots/experimental_grayscale_resistance.pdf}
    \caption{Experimental measurement of resistance (blue) against grayscale (red) plotted on different scales for warming branch and cooling branch. (a) Resistance on log scale against the optically measured grayscale for the warming branch. (b) Resistance on log scale against grayscale for the cooling branch. (c) Resistance curve on linear scale against grayscale for the warming branch.  (d) Resistance on linear scale against grayscale for the cooling branch.}
    \label{fig:resistance_grayscale}
\end{figure}

\begin{figure}[H]
    \centering
    \includegraphics[width=0.4\textwidth]{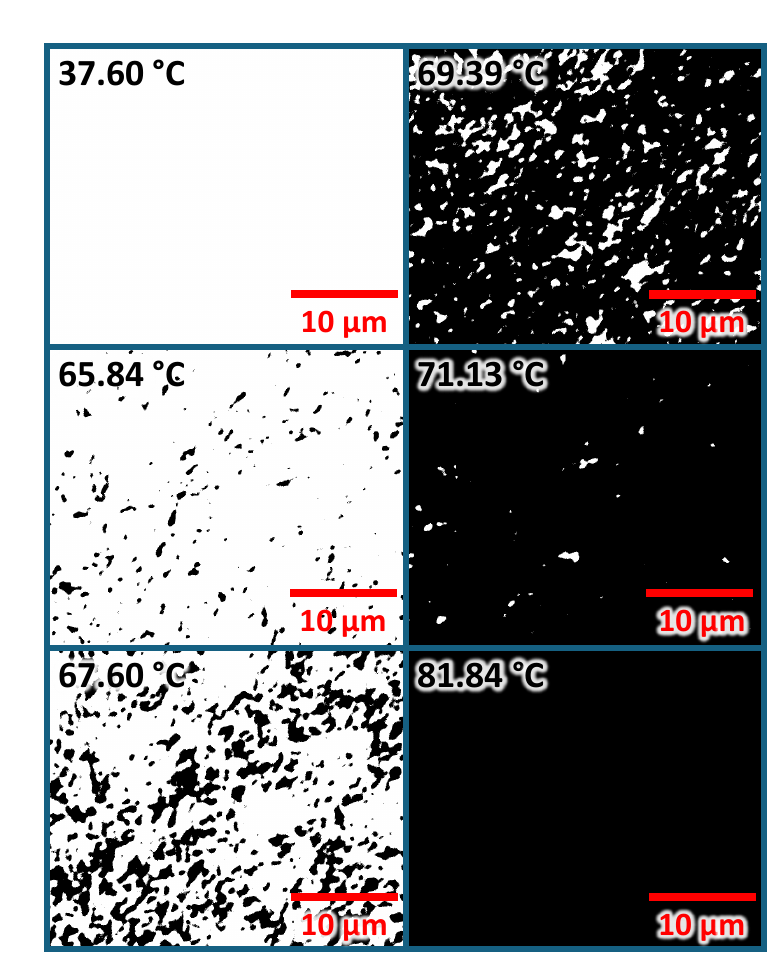}
    \caption{Black and white images of \ce{VO2} using the DOMain INtensity Overturn (DOMINO) method we introduced in~\textcite{Basak2023}. The frames shown are the same as in Fig.~\ref{fig:grayscale_transition_and_fit}.}
    \label{fig:bw_optical_image}
\end{figure}

\section{\label{sec:SI_different_thresholds}Different grayscale thresholds}

\begin{figure}[H]
    \centering
    \includegraphics[width=\textwidth]{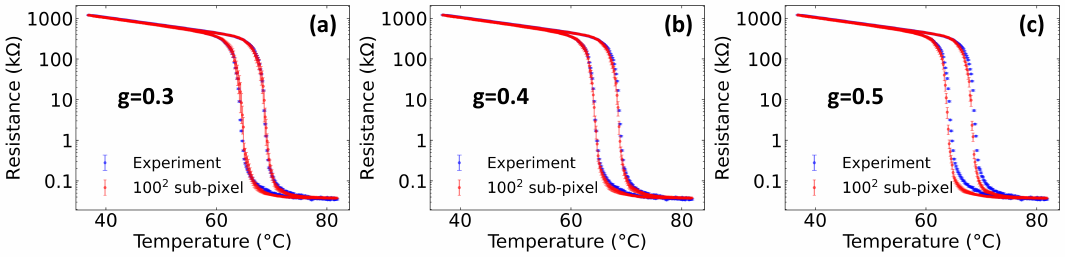}
    \caption{Resistance prediction from $100\times 100$ fractal sub-pixel fits using different grayscale thresholds (a) grayscale threshold set to $g=0.3$. (b) Grayscale threshold $g=0.4$. (c) Grayscale threshold $g=0.5$. Fitting parameters can be found in Table.~\ref{table:fitting-parameters-s100}.}
    \label{fig:grayscale_comparison}
\end{figure}

\section{\label{sec:SI_fitting}Fitting critical 2D RFIM configurations}

\begin{figure}[htbp]
    \centering
    \includegraphics[width=0.5\linewidth]{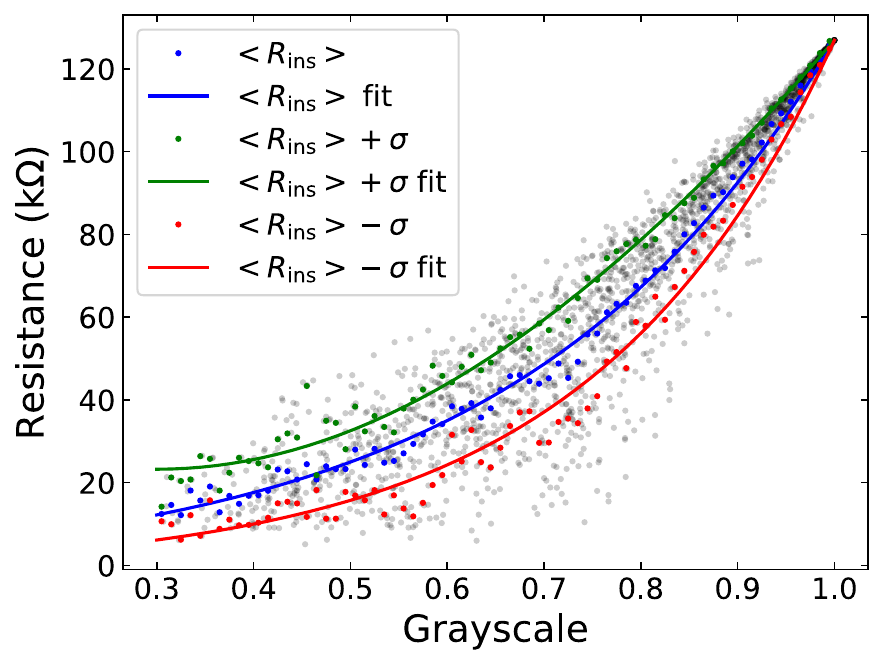}
    \caption{Same figure as Fig.~\ref{fig:sub-pixel_s100_combined}(d), however here the original data points used to determine the phenomenological fits are shown. Green points are determined by finding the point inside a 0.01 bin in grayscale where $17\%$ of data has a greater resistance value than it. Blue points are the mean value within the bin. Red points are the points where $17\%$ of points in the bin have a smaller resistance value than it. Phenomenological fits were determined based on the trend of the data points. For example, the mean values have a curvature at lower grayscale values, however it seems to asymptote to a line at higher grayscale values. Thus, we decided on a square root of a second degree polynomial.}
    \label{fig:fitting_sub-pixel_original}
\end{figure}

\begin{table}[H]
    \centering
    \begin{tabular}{c | c | c | c} 
     \hline
     Branch & Fitting function & Function & Constraint \\
     \hline
     \hline
        Insulating & $f_{\text{sqrt}}(g)$ & $\sqrt{a(g-b)^2 + c} + d$ & $f_{\text{sqrt}}(g=1) = R_{\text{max}}$ \\
        Insulating & $f_{\text{exp}}(g)$ & $ae^{bg} + c$ & $f_{\text{exp}}(g=1) = R_{\text{max}}$ \\
        Insulating & $f_{\text{linear}}(g)$ & $ag + b$ & $f_{\text{linear}}(g=1) = R_{\text{max}}$ \\
        Metallic & $f_{\text{exp}}(g)$ & $ae^{bg} + c$ & $f_{\text{exp}}(g=0) = R_{\text{min}}$ \\
     \hline
    \end{tabular}
    \caption{Functions used for fitting 2D RFIM resistance simulations. $f_{\text{sqrt}}(g)$ is used to fit data for the insulating branch, the function has a constraint where $b$ is the smallest value of grayscale that the function is fit down to, for example in Fig.~\ref{fig:sub-pixel_s100_combined}(d) $b = 0.3$. Additionally, the parameter $d$ is constrained such that $f_{\text{sqrt}}(g=1) = R_{\text{max}}$ where $R_{\text{max}}$ is the maximum value of resistance, thus there are only two free parameters when fitting $f_{\text{sqrt}}(g)$. $f_{\text{exp}}(g)$ is used to fit both insulating and metallic branches, the function has a constraint where $c$ is such that $f_{\text{exp}}(g=1) = R_{\text{max}}$ when fitting the insulating branch and $f_{\text{exp}}(g = 0) = R_{\text{min}}$ when fitting the metallic branch, where $R_{\text{min}}$ is the minimum value of resistance. Thus there are only two free parameters for $f_{\text{exp}}$. $f_{\text{linear}}(g)$ is used to fit data for the insulating branch, $b$ is constrained such that when $f(g = 1) = R_{\text{max}}$, thus there is only one free parameter. Fitting form is the function used for fitting after applying the constraints. Values and error bars for these fit parameters are reported in Table.~\ref{table:fitting-parameters-s10} and Table.~\ref{table:fitting-parameters-s100}.}
    \label{table:fitting-functions}
\end{table}

We can derive the value of the constrained parameter in terms of the fitting parameters like so:

\begin{enumerate}[label = (\alph*)]
    \item Insulating $f_{\rm sqrt}(g)$
    \begin{align*}
        f(g = 1) &= \sqrt{a(1 - b)^2 + c} + d = R_{\rm max}\\
        d &= R_{\rm max} - \sqrt{a(1 - b)^2 + c}
    \end{align*}
    \item Insulating $f_{\rm exp}(g)$
    \begin{align*}
        f(g=1) &= ae^b + c = R_{\rm max}\\
        c &= R_{\rm max} - ae^b
    \end{align*}
    \item Insulating $f_{\rm linear}(g)$
    \begin{align*}
        f(g = 1) &= a + b = R_{\rm max}\\
        b &= R_{\rm max} - a
    \end{align*}
    \item Metallic $f_{\rm exp}(g)$
    \begin{align*}
        f(g = 0) &= a + c = R_{\rm min}\\
        c &= R_{\rm min} - a
    \end{align*}
\end{enumerate}

\begin{table}[H]
    \centering
    \begin{tabular}{l | l | l}
    \hline
        Curve & Function & Parameters\\
        \hline
        \hline
        $\braket{R_{\text{ins}}}+\sigma$ & $f_{\text{linear}}(g)$ & $a = 1.42\mathrm{E}6\pm 1.95\mathrm{E}4$\\
        \hline
        ${\braket{R_{\text{ins}}}}$ & ${f_{\text{exp}}(g)}$ & ${a = 9.78\mathrm{E}4 \pm 1.13\mathrm{E}3, b = 2.48\pm 9.52\mathrm{E}-3}$\\
        \hline
        $\braket{R_{\text{ins}}}-\sigma$ & $f_{\text{exp}}(g)$ & $a = 1.28\mathrm{E}4 \pm 1.75\mathrm{E}2, b = 4.41\pm 1.27\mathrm{E}-3$\\
        \hline
        $\braket{R_{\text{met}}}+\sigma$ & $f_{\text{exp}}(g)$ & $a = 1.59\mathrm{E}5\pm 1.20\mathrm{E}4, b = 1.73\mathrm{E}-3\pm 1.31\mathrm{E}-4$ \\
        \hline
        ${\braket{R_{\text{met}}}}$ & ${f_{\text{exp}}(g)}$ & ${a = 1.17\mathrm{E}5 \pm 4.51\mathrm{E}3, b = 1.58\mathrm{E}-3\pm 6.09\mathrm{E}-5}$\\
        \hline
        $\braket{R_{\text{met}}}-\sigma$ & $f_{\text{exp}}(g)$ & $a = 6.30\mathrm{E}4\pm 1.53\mathrm{E}3, b = 1.46\mathrm{E}-3\pm 3.56\mathrm{E}-5$\\
        \hline
    \end{tabular}
    \caption{Parameters for fitting functions used in 2D RFIM resistance simulation plots with sub-pixel size $10\times 10$. The errors for parameters were calculated by setting one free parameter and fixing all others at the optimal value. If a grayscale threshold $g=0.32$ is chosen, then the insulating branch is fit down to the nearest tenth less than the threshold, in this case $g=0.3$. The metallic branch is fit up to the nearest tenth greater than the threshold, in this case $g=0.4$.}
    \label{table:fitting-parameters-s10}
\end{table}

\begin{table}[H]
    \centering
    \begin{tabular}{ l | l | l | l}
    \hline
    Figure & Curve & Function & Parameters\\
        \hline
        \hline
        Figure.~\ref{fig:grayscale_comparison}(a) & $\braket{R_{\text{ins}}}+\sigma$ & $f_{\text{sqrt}}(g)$ & $a = 2.26\mathrm{E}13\pm 1.34\mathrm{E}11, b = 0.3, c = 4.80\mathrm{E}-2 \pm 3.23\mathrm{E}-4$\\
        \hline
        & $\braket{R_{\text{ins}}}$ & ${f_{\text{sqrt}}(g)}$ & ${a = 3.10\mathrm{E}16\pm 1.44\mathrm{E}14, b=0.3, c= 6.67\mathrm{E}1\pm 3.10\mathrm{E}-1}$\\
        \hline
        & $\braket{R_{\text{ins}}}-\sigma$ & $f_{\text{exp}}(g)$ & $a = 2.35\mathrm{E}4 \pm 1.33\mathrm{E}2, b = 4.00\pm 5.16\mathrm{E}-3$\\
        \hline
        & $\braket{R_{\text{met}}}+\sigma$ & $f_{\text{exp}}(g)$ & $a = 2.27\mathrm{E}1 \pm 6.21\mathrm{E}-1, b = 6.17 \pm 8.61\mathrm{E}-2$\\
        \hline
         & ${\braket{R_{\text{met}}}}$ & ${f_{\text{exp}}(g)}$ & ${a = 2.40\mathrm{E}1 \pm 4.01\mathrm{E}-1, b = 4.93\mathrm{E}-2}$\\
        \hline
        & $\braket{R_{\text{met}}}-\sigma$ & $f_{\text{exp}}(g)$ & $a = 4.31\mathrm{E}1 \pm 4.46\mathrm{E}-1, b = 2.68\pm 2.05\mathrm{E}-2$\\
        \hline
        Figure.~\ref{fig:combined_results}(c) & $\braket{R_{\text{ins}}}+\sigma$ & $f_{\text{sqrt}}(g)$ & $a = 2.26\mathrm{E}13\pm 1.34\mathrm{E}11, b= 0.3,  c = 4.80\mathrm{E}-2 \pm 3.23\mathrm{E}-4$\\
        \hline
        & $\bm{\braket{R_{\text{ins}}}}$ & $\bm{f_{\text{sqrt}}(g)}$ & $\bm{a = 3.10\mathrm{E}16\pm 1.44\mathrm{E}14, b=0.3, c= 6.67\mathrm{E}1\pm 3.10\mathrm{E}-1}$\\
        \hline
        & $\braket{R_{\text{ins}}}-\sigma$ & $f_{\text{exp}}(g)$ & $a = 2.35\mathrm{E}4 \pm 1.33\mathrm{E}2, b = 4.00\pm 5.16\mathrm{E}-3$\\
        \hline
        & $\braket{R_{\text{met}}}+\sigma$ & $f_{\text{exp}}(g)$ & $a = 3.28\mathrm{E}1 \pm 9.69\mathrm{E}-1, b = 5.24\pm 7.28\mathrm{E}-2$\\
        \hline
        & $\bm{\braket{R_{\text{met}}}}$ & $\bm{f_{\text{exp}}(g)}$ & $\bm{a = 3.59\mathrm{E}1 \pm 5.73\mathrm{E}-1, b = 4.17 \pm 3.63\mathrm{E}-2}$\\
        \hline
        & $\braket{R_{\text{met}}}-\sigma$ & $f_{\text{exp}}(g)$ & $a = 4.41\mathrm{E}1 \pm 6.16\mathrm{E}-1, b = 2.66 \pm 2.50\mathrm{E}-2$\\
        \hline
        Figure.~\ref{fig:grayscale_comparison}(b) & $\braket{R_{\text{ins}}}+\sigma$ & $f_{\text{sqrt}}(g)$ & $a = 1.14\mathrm{E}13\pm 8.79\mathrm{E}10,b=0.4,  c = 1.64\mathrm{E}-2 \pm 1.71\mathrm{E}-4$\\
        \hline
        & ${\braket{R_{\text{ins}}}}$ & ${f_{\text{sqrt}}(g)}$ & ${a= 1.30\mathrm{E}16\pm 4.79\mathrm{E}13, b=0.4, c = 2.21\mathrm{E}1 \pm 8.15\mathrm{E}-2}$\\
        \hline
        & $\braket{R_{\text{ins}}}-\sigma$ & $f_{\text{sqrt}}(g)$ & $a = 3.65\mathrm{E}16\pm 2.58\mathrm{E}14, b=0.4, c = 5.52\mathrm{E}1 \pm 3.91\mathrm{E}-1$\\
        \hline
        & $\braket{R_{\text{met}}}+\sigma$ & $f_{\text{exp}}(g)$ & $a = 3.28\mathrm{E}1 \pm 9.69\mathrm{E}-1, b = 5.24\pm 7.28\mathrm{E}-2$\\
        \hline
        & ${\braket{R_{\text{met}}}}$ & ${f_{\text{exp}}(g)}$ & ${a = 3.59\mathrm{E}1 \pm 5.73\mathrm{E}-1, b = 4.17 \pm 3.63\mathrm{E}-2}$\\
        \hline
        & $\braket{R_{\text{met}}}-\sigma$ & $f_{\text{exp}}(g)$ &$a = 4.41\mathrm{E}1 \pm 6.16\mathrm{E}-1, b = 2.66 \pm 2.50\mathrm{E}-2$\\
        \hline
        Figure.~\ref{fig:grayscale_comparison}(c) & $\braket{R_{\text{ins}}}+\sigma$ & $f_{\text{sqrt}}(g)$ & $a = 7.78\mathrm{E}12\pm 6.64\mathrm{E}10, b=0.5, c = 6.41\mathrm{E}-3\pm 1.04\mathrm{E}-4$\\
        \hline
        & ${\braket{R_{\text{ins}}}}$ & ${f_{\text{sqrt}}(g)}$ & ${a = 2.25\mathrm{E}13\pm 1.22\mathrm{E}11, b=0.5, c = 2.35\mathrm{E}-2 \pm 1.59\mathrm{E}-4}$\\
        \hline
        & $\braket{R_{\text{ins}}}-\sigma$ & $f_{\text{sqrt}}(g)$ & $a = 9.35\mathrm{E}13\pm 6.59\mathrm{E}11, b=0.5, c=9.99\mathrm{E}-2 \pm 7.50\mathrm{E}-4$\\
        \hline
        & $\braket{R_{\text{met}}}+\sigma$ & $f_{\text{exp}}(g)$ & $a = 3.55\mathrm{E}1 \pm 7.766\mathrm{E}-1, b = 4.91 \pm 4.49\mathrm{E}-2$\\
        \hline
        & ${\braket{R_{\text{met}}}}$ & ${f_{\text{exp}}(g)}$ & ${a = 3.63\mathrm{E}1 \pm 4.64\mathrm{E}-1, b = 4.10 \pm 2.50\mathrm{E}-2}$\\
        \hline
        & $\braket{R_{\text{met}}}-\sigma$ & $f_{\text{exp}}(g)$ & $a = 4.10\mathrm{E}1 \pm 7.40\mathrm{E}-1, b = 2.82 \pm 3.03\mathrm{E}-2$\\
        \hline
    \end{tabular}
    \caption{Parameters for fitting functions used in 2D RFIM resistance simulation plots with fractal sub-pixel size $100\times 100$. The errors for parameters were calculated by setting one free parameter and fixing all others at the optimal value. If a grayscale threshold $g=0.32$ is chosen, then the insulating branch is fit down to the nearest tenth less than the threshold, in this case $g=0.3$. The metallic branch is fit up to the nearest tenth greater than the threshold, in this case $g=0.4$. The fit curves used for the best prediction in Fig.~\ref{fig:combined_results}(c) are highlighted in bold.}
    \label{table:fitting-parameters-s100}
\end{table}

\end{document}